\documentclass[a4paper,11pt]{article}
\pdfoutput=1

\usepackage{a4}
\usepackage{jheppub}
\usepackage{graphics}
\usepackage{amssymb,amsmath}
\usepackage{color}
\usepackage{wasysym}
\usepackage[tight]{subfigure}
\usepackage{booktabs}
\usepackage{multirow}
\usepackage{rotating}
\usepackage{float}
\usepackage{diagbox}
\usepackage{ulem}

\newcommand{\beq}{\begin{equation}}
\newcommand{\eeq}{\end{equation}}
\newcommand{\beqn}{\begin{eqnarray}}
\newcommand{\eeqn}{\end{eqnarray}}

\newcommand{\mneu}[1]{m_{\tilde{\chi}^0_{#1}}}
\newcommand{\mcha}[1]{m_{\tilde{\chi}^\pm_{#1}}}

\newcommand{\neu}[1]{\tilde{\chi}^0_{#1}}
\newcommand{\cha}[1]{\tilde{\chi}^\pm_{#1}}

\newcommand{\gev}{\ \mathrm{GeV}}
\newcommand{\tev}{\ \mathrm{TeV}}

\newcommand{\met}{E_T^\mathrm{miss}}

\newcommand{\tilW}{\widetilde W}
\newcommand{\tilH}{\widetilde H}

\newcommand{\BR}{ {\rm BR} }

\def\gsim  {\hspace{0.3em}\raisebox{0.4ex}{$>$}\hspace{-0.75em}\raisebox{-.7ex}{$\sim$}\hspace{0.3em}}
\def\lsim  {\hspace{0.3em}\raisebox{0.4ex}{$<$}\hspace{-0.75em}\raisebox{-.7ex}{$\sim$}\hspace{0.3em}}

\preprint{\parbox{4cm}{KCL-PH-TH/2014-39\\LCTS/2014-38}}
\title{Prospects for observing charginos and neutralinos at a 100 TeV proton-proton collider}
\author[a,b]{Bobby S. Acharya,}
\author[a]{Krzysztof Bo\.{z}ek,}
\author[a]{Chakrit Pongkitivanichkul,}
\author[a]{Kazuki Sakurai}
\affiliation[a]{Theoretical Particle Physics \&
  Cosmology, Department of Physics, King's College London, London, WC2R 2LS, United Kingdom}
\affiliation[b]{The Abdus Salam International Centre for Theoretical Physics, Trieste, Italy}
\emailAdd{bobby.acharya@kcl.ac.uk}
\emailAdd{krzysztof.bozek@kcl.ac.uk}
\emailAdd{chakrit.pongkitivanichkul@kcl.ac.uk}
\emailAdd{kazuki.sakurai@kcl.ac.uk}
\abstract{
We investigate the prospects for discovering charginos and neutralinos at a future $pp$ collider with $\sqrt{s} = 100$ TeV.
We focus on models where squarks and sleptons are decoupled -- as motivated by the LHC data.
Our initial study is based on models where Higgsinos form the main component of the LSP and $W$-inos compose the heavier chargino states ($M_2 > \mu$), though
it is straightforward to consider the reverse situation also.   
We show that in such scenarios $W$-inos decay into $W^\pm$, $Z$ and $h$ plus neutralinos almost universally.
In the $W Z$ channel we compare signal and background in various kinematical distributions.  
We design simple but effective signal regions for the trilepton channel and evaluate discovery reach and exclusion limits. 
Assuming 3000 fb$^{-1}$ of integrated luminosity, $W$-inos could be discovered (excluded) up to 1.1 (1.8) TeV if the spectrum is not compressed.
\color{black}
}
\keywords{Supersymmetry, Hadron-Hadron Scattering}

\begin{document}
\maketitle
\flushbottom

%%%%%%%%%%%%%%%%%%%%%%%%%%%%%%%%%%%%%%%%%%%%%%%%%%%%%%%%%%%%%%%%%%%%%%%%%%%%%%%%%%
%%%%%%%%%%%%%%%%%%%%%%%%%%%%%%%%%%%%%%%%%%%%%%%%%%%%%%%%%%%%%%%%%%%%%%%%%%%%%%%%%%

\section{Introduction \label{sec:intro}}

Supersymmetry is one of the most promising ideas for physics beyond the Standard Model, since it directly addresses the hierarchy problem.
In its minimal incarnation, the MSSM, the chargino-neutralino sector is particularly important for several phenomenological reasons.
Firstly, this sector contains Higgsinos, whose mass parameter, $\mu$, plays a crucial role in electroweak symmetry breaking.
If the MSSM provides a solution to the gauge hierarchy problem, at least some of the charginos and neutralinos must be present not too far from the electroweak scale.
Secondly, many SUSY breaking scenarios suggest that one of the neutralinos becomes the lightest SUSY particle (LSP).
Typically, the lightest neutralino is stable due to a discrete symmetry (e.g. R-parity) and might be a promising candidate for dark matter.
Such a stable neutralino also plays a crucial role in the collider phenomenology since the decay of supersymmetric particles will always 
produce the LSP, leading to a distinctive missing energy signature.

The ATLAS and CMS experiments at the CERN Large Hadron Collider (LHC) have put considerable effort into looking for charginos and neutralinos in the LHC data.
In hadron colliders the expected limit and discovery reach for the charginos and neutralinos are considerably weaker compared to those for squarks and gluinos.
For the $\cha{1} \neu{2} \to W^\pm \neu{1} Z \neu{1}$ simplified model with $\mcha{1} = \mneu{2}$ and $\mneu{1} = 0$ GeV, the current limit is 
$\mcha{1} \gsim 400$ GeV \cite{Aad:2014vma, Khachatryan:2014qwa}.
The projection for the 14 TeV LHC has been estimated for the same simplified model by ATLAS \cite{ATLAS:2014projection}.
The 5-$\sigma$ discovery reach (95\% CL limit) for the chargino mass is about 550 (880) GeV for 300 fb$^{-1}$ and 800 (1100) GeV for 3000 fb$^{-1}$.
%as summarised in Table \ref{tab:projection}.  
For massive neutralinos ($\mneu{1} > 0$ GeV) or models with $\BR(\neu{2} \to h \neu{1}) > 0$, the limit and discovery reach become even weaker. These limits are well below those required by typical dark matter model.

%\begin{table}[t!]
%\begin{center}
%\begin{tabular}{|c|c|c|c|}
%\hline
%Energy   &   Luminosity   &  5-$\sigma$ discovery   &   95\% CL limit   \\
%\hline
%\hline 
%8 TeV    &   $\sim 20$ fb$^{-1}$   &         ---          &  $\sim 400$ GeV     \\
%\hline
%\hline 
%14 TeV  &   300 fb$^{-1}$       &   $\sim 550$ GeV       &   $\sim 800$ GeV   \\
%\hline 
%14 TeV  &   3000 fb$^{-1}$       &   $\sim 800$ GeV       &   $\sim 1100$ GeV   \\
%\hline
%\end{tabular}
%\caption{
%The 5-$\sigma$ discovery reach and 95\% CL limit on the chargino mass ($\mcha{1} = \mneu{2}$) at the LHC, assuming 
%the $WZ$ channel in the $\cha{1} \neu{2} \to W^\pm \neu{1} Z \neu{1}$ topology with massless neutralino ($\mneu{1} = 0$ GeV). 
%The first row shows the current limit obtained by ATLAS and CMS experiments \cite{Aad:2014vma, Khachatryan:2014qwa}, and the second and third row represent the projected 
%values derived by ATLAS \cite{ATLAS:2014projection}. 
%\label{tab:projection}
%}
%\end{center}
%\end{table}%

Recently, there has been discussion of the next generation of circular colliders, including high energy proton-proton machines.
Several physics cases at proton-proton colliders with $\sqrt{s} \simeq 100$ TeV have already been studied \cite{Cohen:2013xda, Andeen:2013zca, Apanasevich:2013cta, Stolarski:2013msa, Yu:2013wta, Zhou:2013raa, Jung:2013zya, Low:2014cba, Cirelli:2014dsa}. 
In particular, the limit and discovery reach for coloured SUSY particles have been studied in the context of simplified models 
assuming a 100 TeV proton-proton collider with $3000$ fb$^{-1}$ of integrated luminosity \cite{Cohen:2013xda}.
The mono-jet search \cite{Zhou:2013raa} as well as the mono-photon, soft lepton and disappearing track searches \cite{Low:2014cba, Cirelli:2014dsa} have been studied in the similar setup
for production of the pure $W$-inos (Higgsinos), assuming they are the main component of the LSP. 
%The mono-jet search, soft lepton search  for the lightest neutralino pair production has also been studied in the similar setup \cite{Zhou:2013raa}.
The 100 TeV colliders will provide a great opportunity to discover heavier charginos and neutralinos beyond the LHC reach.

In this paper we study chargino-neutralino search at a 100 TeV collider assuming 3000 (1000) fb$^{-1}$ luminosity exploiting the $WZ$ channel.
In stead of employing a simplified model approach, we work on a model which may arise as a limit of concrete models.
In particular we assume $M_2 > \mu > 0$ and $M_2 - \mu \gg m_Z$, where $M_2$ is the $W$-ino mass and $\mu$ is the Higgsino mass. 
In this scenario Higgsinos form the main component of the lighter charginos and neutralinos ($\cha{1}, \neu{1}, \neu{2} \sim \tilH^\pm, \tilH^0_1, \tilH^0_2$) and $W$-inos compose the heavier charginos and neutralinos ($\cha{1}, \neu{3} \sim \tilW^\pm, \tilW^0$).    
This assumption is partly motivated by naturalness, by anomaly mediation SUSY breaking scenarios, by string/$M$ theory models and by split supersymmetry \cite{Randall:1998uk, Giudice:1998xp, Moroi:1999zb, Acharya:2007rc, Acharya:2008zi, Acharya:2012tw, ArkaniHamed:2004fb, Giudice:2004tc, ArkaniHamed:2004yi}\\
\indent The rest of the paper is organised as follows.  In section \ref{sec:xsec-br}, we describe the model setup and study the production cross sections and branching ratios of charginos and neutralinos.  
After clarifying our simulation setup in section \ref{sec:sim}, various kinematic distributions for signal and background are studied in section \ref{sec:distributions}, which will be used to design optimal event selection cuts for the chargino-neutralino search.
In section \ref{sec:limit}, we present the result of our analysis and derive the limit and discovery reach in the $M_2 - \mu$ parameter plane.
The conclusions are given in section \ref{sec:concl}.

\section{The cross sections and branching ratios \label{sec:xsec-br}}

\subsection{The model setup \label{sec:model}}

In this paper we focus on the models with $M_2 > \mu > 0$ and $M_2 - \mu \gg m_Z$, where the $\mu$ is the mass of the Higgsinos and $M_2$ is the mass of the $W$-inos since the $W$-ino production cross-section is larger than the Higgsino cross-section.
We assume that all the other SUSY particles, including the $B$-ino, are decoupled and all SUSY breaking parameters are real for simplicity.
In this situation the mixing between $W$-ino and Higgsino is negligible;
the two Higgsino doublets are the lightest charginos and the two lightest neutralinos (which are almost degenerate) and the $W$-inos (SU(2) triplet) are the second lightest charginos and the third lightest neutralino (almost mass degenerate):
\beqn
\cha{1}, \neu{1}, \neu{2} &\sim& \tilH^\pm, \tilH^0_1, \tilH^0_2 ~~~~{\rm with}~~\mcha{1} \simeq \mneu{1} \simeq \mneu{2} \simeq |\mu|,  \nonumber \\
\cha{2}, \neu{3} &\sim& \tilW^\pm, \tilW^0 ~~~\,~~~~~{\rm with}~~\mcha{2} \simeq \mneu{3} \simeq |M_2|,
\eeqn
where $\tilH_{1/2}^0 = \frac{1}{\sqrt{2}} (\tilH_u^0 \mp \tilH_d^0)$ is the neutral Higgsino mass eigenstate.   
With this setup, the remaining free parameters are $M_2$, $\mu$ and $\tan\beta$.
We use $\tan \beta = 10$ throughout our numerical study.
However, the impact of $\tan \beta$ on the production cross section and branching ratio of the charginos and neutralinos that are $W$-ino or Higgsino like
is almost negligible unless $\tan \beta$ is extremely small.  
We therefore believe our results including the chargino-neutrino mass reach are still useful for other values of $\tan\beta$.

\subsection{The cross sections \label{sec:xsec}}

\begin{figure}
	\centering \vspace{-0.0cm}
		\includegraphics[width=0.7\textwidth]{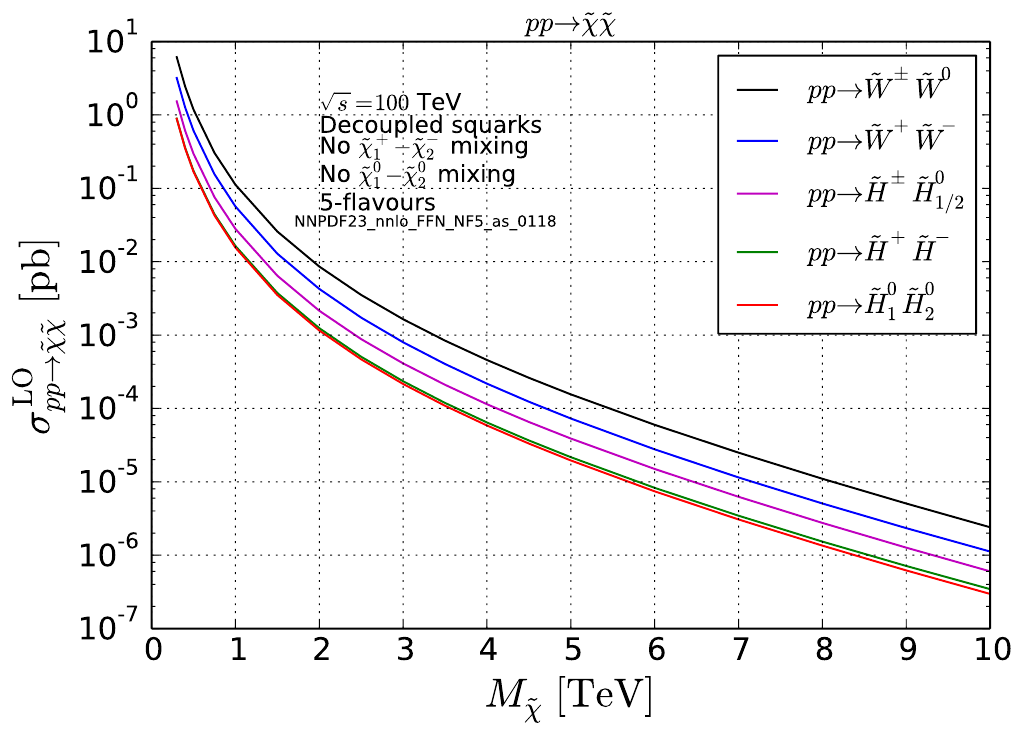}
	\caption{ 
	The leading order cross sections for the $W$-ino and Higgsino pair productions at a 100 TeV proton-proton collider with decoupled squarks and sleptons.
\label{fig:prod-xsec}}
\end{figure}

We show the leading order (LO) cross sections for the $W$-ino and Higgsino pair productions at a 100 TeV proton-proton collider in Fig.~\ref{fig:prod-xsec}.
The cross sections are calculated using {\tt MadGraph\,5} \cite{Alwall:2011uj}.
Since squarks are decoupled, the $W$-inos and Higgsinos are produced via the $s$-channel diagrams exchanging off-shell $W^\pm$ and $Z$ bosons.
For the pure $W$-inos and Higgsinos, there is no associated $W$-ino-Higgsino production process.
Pair production of the same neutralino states, $\tilW^0 \tilW^0$, $\tilH_1^0 \tilH_1^0$, $\tilH_2^0 \tilH_2^0$, are also absent.

One can see that the $\tilW^\pm \tilW^0$ production mode has the largest cross section.
The LO cross section varies from $10^{3}$  fb to $10^{-2}$ fb for the $W$-ino mass from 500 GeV to 8 TeV.
%Therefore, 100 TeV colliders with 3 ab$^{-1}$ can produce 30 $W$-ino pairs with the 8 TeV mass.  

\subsection{The branching ratios \label{sec:br}}

%With the assumption of $M_2 > \mu$ and decoupled squarks and sleptons, we are interested in the $W$-ino production and their decay into Higgsino states.  
The $W$-ino-Higgsino interaction is derived from the kinetic terms of Higgsinos.
\beqn
{\cal L} &\supset& \Big[ H_u^\dagger e^V H_u + H_d^\dagger e^V H_d  \Big]_{\theta^4}  \nonumber \\
&\supset& \sqrt{2} g ( H_u^* \tilW^a T^a \tilH_u - H_d^* \tilW^a T^a \tilH_d) + {\rm h.c.}  ~.
\label{eq:interaction}
\eeqn 
The Higgs and Higgsino fields can be written in terms of the Goldstone bosons and the mass eigenstates as:
\beq
\begin{pmatrix}
H_u^+ \\ H_u^0
\end{pmatrix}
=
\begin{pmatrix}
\sin\beta \cdot \phi^+ + \cdots  \\ 
\frac{1}{\sqrt{2}} ( \cos\alpha \cdot h + i \sin\beta \cdot \phi^0 ) + \cdots
\end{pmatrix}, ~~~
\begin{pmatrix}
\tilH_u^+ \\ \tilH_u^0
\end{pmatrix}
\simeq
\begin{pmatrix}
\tilH^+ \\ \frac{1}{\sqrt{2}} (\tilH_1^0 + i \tilH_2^0)
\end{pmatrix},~~~\,
\nonumber
\eeq
\beq
~~
\begin{pmatrix}
H_d^0 \\ H_d^-
\end{pmatrix}
=
\begin{pmatrix}
\frac{- 1}{\sqrt{2}} ( \sin\alpha \cdot h + i \cos\beta \cdot \phi^0 ) + \cdots \\
- \cos\beta \cdot \phi^- + \cdots   
\end{pmatrix}, ~~~
\begin{pmatrix}
\tilH_d^0 \\ \tilH_d^-
\end{pmatrix}
\simeq
\begin{pmatrix}
\frac{1}{\sqrt{2}} (\tilH_1^0 - i \tilH_2^0) \\ \tilH^-
\end{pmatrix},
\label{eq:higgs}
\eeq
where $h$ is the SM like Higgs boson, and $\phi^0$ and $\phi^\pm$ are the Goldstone bosons to be eaten by the SM gauge bosons, $Z$ and $W^\pm$, respectively. 
The angles $\alpha$ and $\beta$ represent the mixing for the neutral and charged Higgs mass matrices.

In the large $\tan\beta$ limit, we have $\cos\alpha/\sin\alpha \simeq (-\sin\beta)/\cos\beta$, and one can see that 
the $h \tilW \tilH$, $\phi^0 \tilW \tilH$ and $\phi^\pm \tilW \tilH$ have the same coupling. 
In this limit one can find the following results using the Goldstone equivalence theorem \cite{Jung:2014bda}.
\beqn
\BR(\tilW^\pm) &\simeq& \left\{ \begin{array}{ll}
0.5~~~ & \to W^\pm \tilH_1^0 ~{\rm or}~ W^\pm  \tilH_2^0   \\
0.25~~ & \to h \tilH^\pm   \\
0.25~~ & \to Z \tilH^\pm   \\
\end{array} \right.
\nonumber 
\\
%\eeq
%\beq
%~~~
\BR(\tilW^0) &\simeq& \left\{ \begin{array}{ll}
0.5~~~ & \to W^\pm \tilH^\mp  \\ 
0.25~~ & \to h \tilH^0_1 ~{\rm or}~ h \tilH^0_2  \\
0.25~~ & \to Z \tilH^0_1 ~{\rm or}~ Z \tilH^0_2   \\
\end{array} \right..
%\nonumber \\
\label{eq:br}
\eeqn
The different CP properties between $h$ and $\phi^0$, and $\tilH_1^0$ and $\tilH_2^0$ result in the
different rates for $\tilW^0 \to h \tilH^0_1$ and $\tilW^0 \to Z \tilH_1^0, h \tilH_2^0$.  
These rates are given by
\beqn
\BR(\tilW^\pm \to W^\pm \tilH_1^0) &\simeq& \BR(\tilW^\pm \to W^\pm \tilH_2^0), \nonumber \\
\BR(\tilW^0 \to h \tilH_{1/2}^0) &\simeq& \BR(\tilW^0 \to Z \tilH_{2/1}^0), \nonumber \\
\frac{\BR(\tilW^0 \to Z \tilH_1^0)}{\BR(\tilW^0 \to h \tilH_1^0)} &\simeq& \frac{1 - 2 |\mu / M_2| }{ 1 + 2 |\mu/M_2|} ~.   
\eeqn

\begin{figure}[t]
\begin{center}
  \subfigure[]{\includegraphics[width=0.48\textwidth]{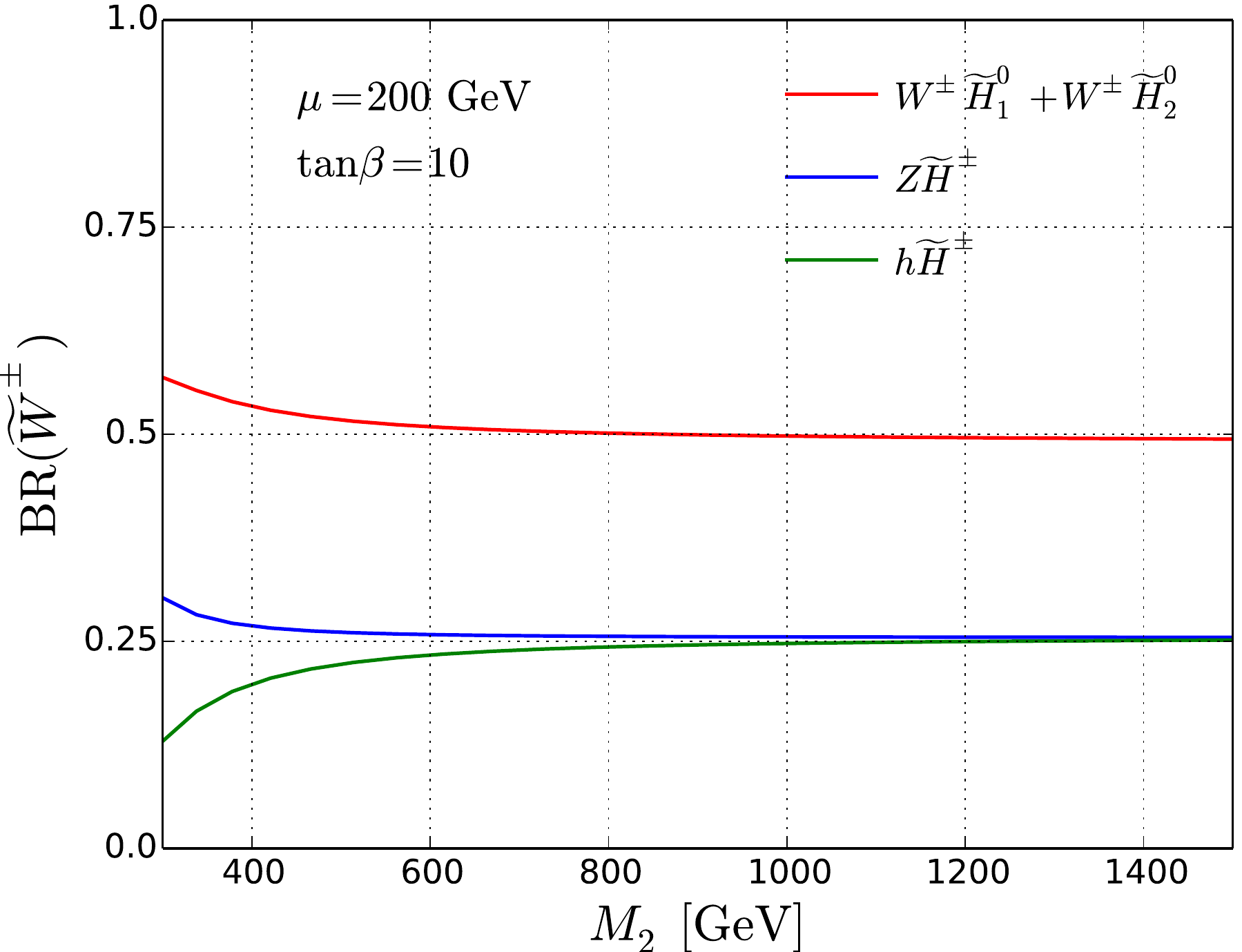}\label{fig:BR-cha2}} \hspace{0.2cm}
  \subfigure[]{\includegraphics[width=0.48\textwidth]{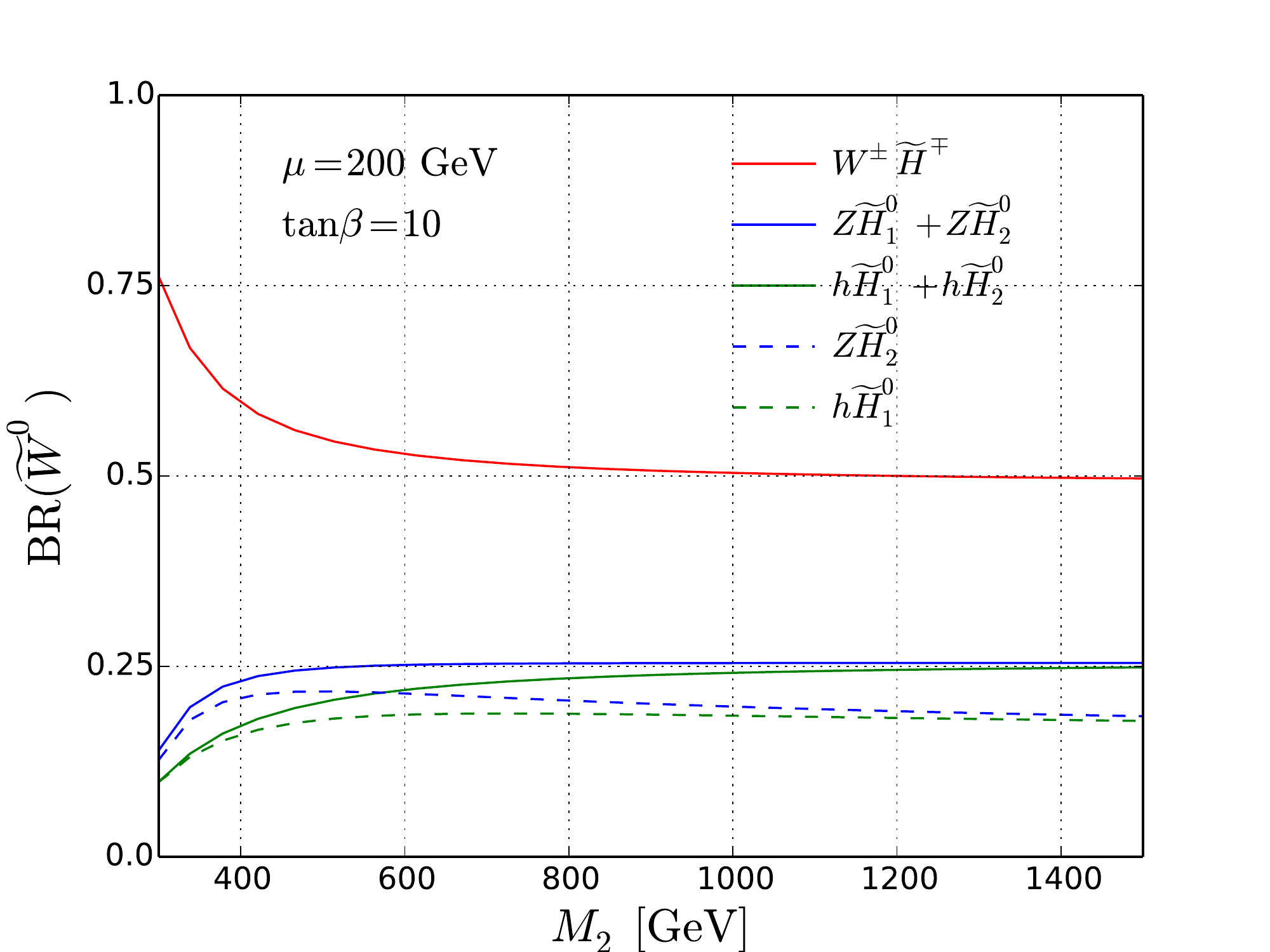}\label{fig:BR-neu3}} 
\caption{  The branching ratios of $\tilW^\pm$ {\bf (a)} and $\tilW^0$ {\bf (b)} as functions of $M_2$.   
The $\mu$ parameter is fixed at $200$~GeV.
The SUSY particles other than $W$-inos and Higgsinos are decoupled. 
 \label{fig:br} }
\end{center}
\end{figure}

\begin{figure}[t!]
	\centering \vspace{-0.0cm}
		\includegraphics[width=0.45\textwidth]{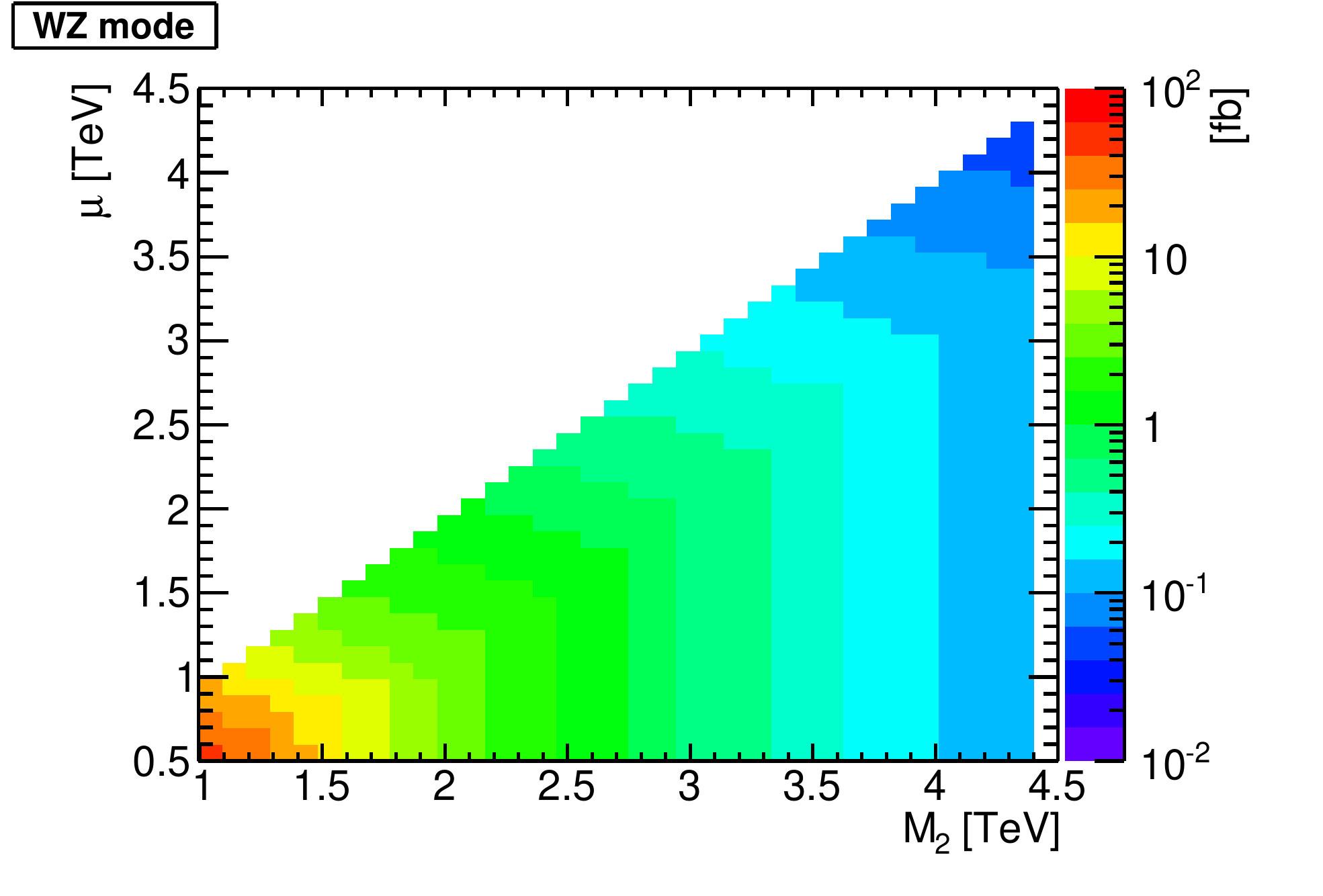}
		\includegraphics[width=0.45\textwidth]{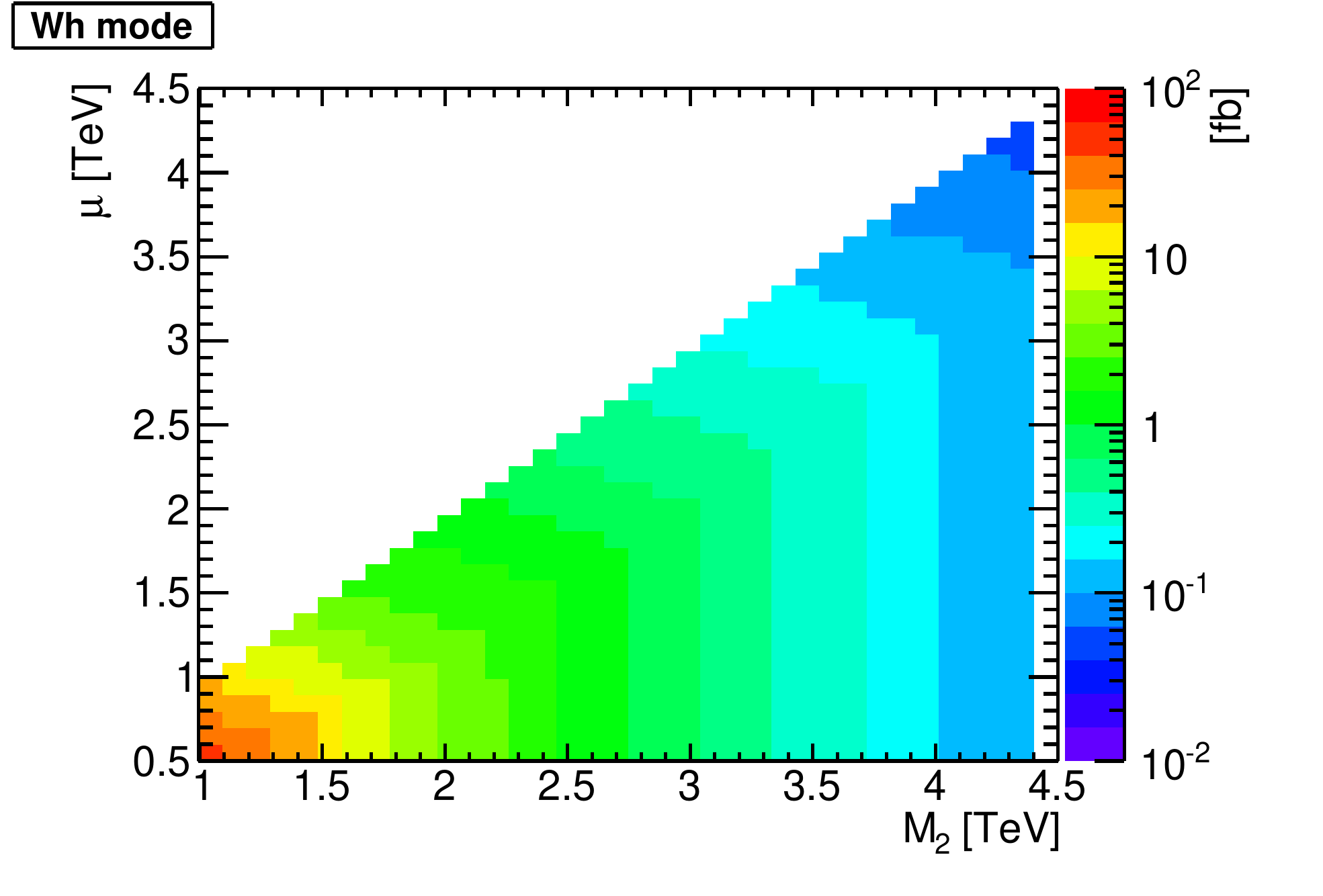}
		\includegraphics[width=0.45\textwidth]{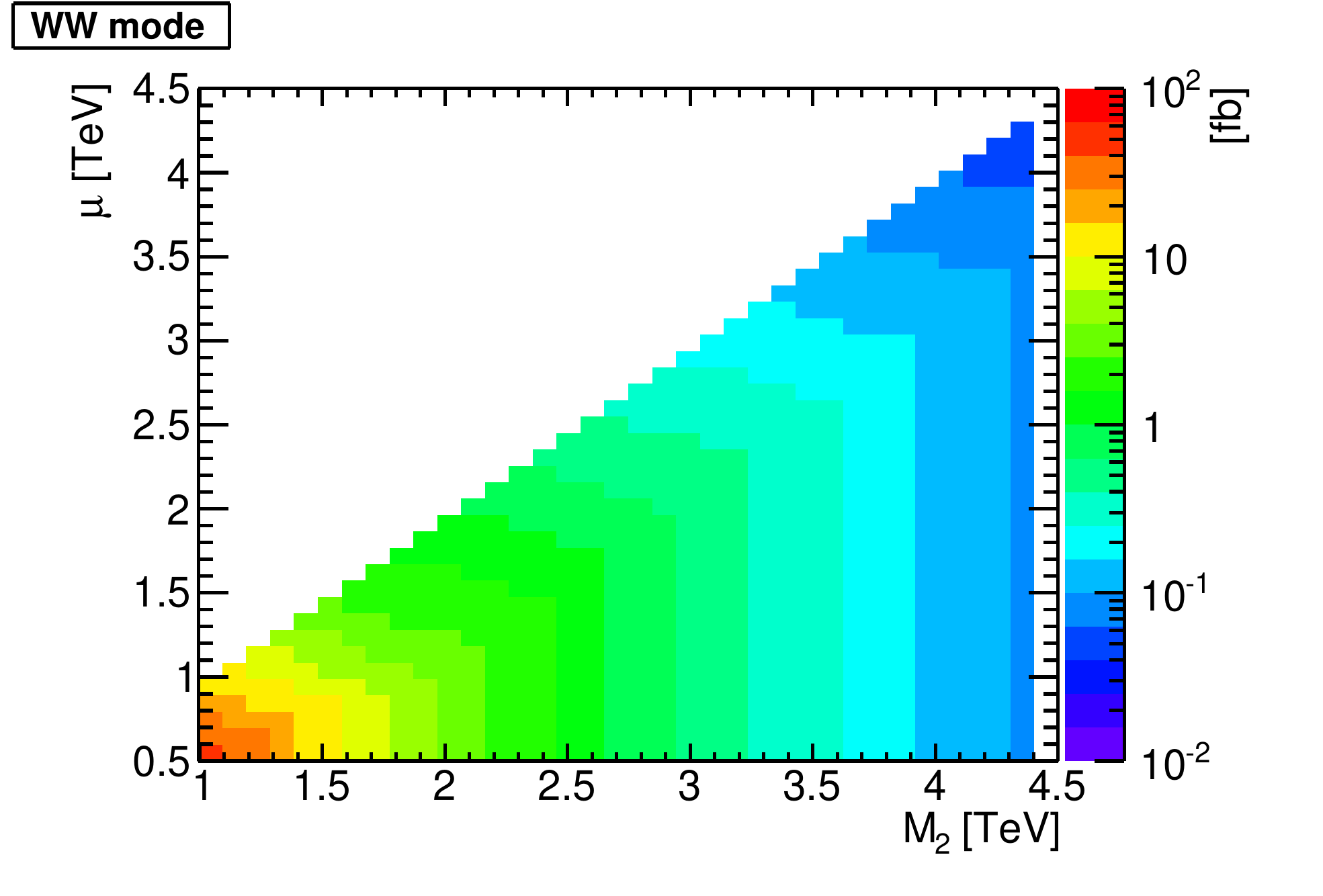}		
		\includegraphics[width=0.45\textwidth]{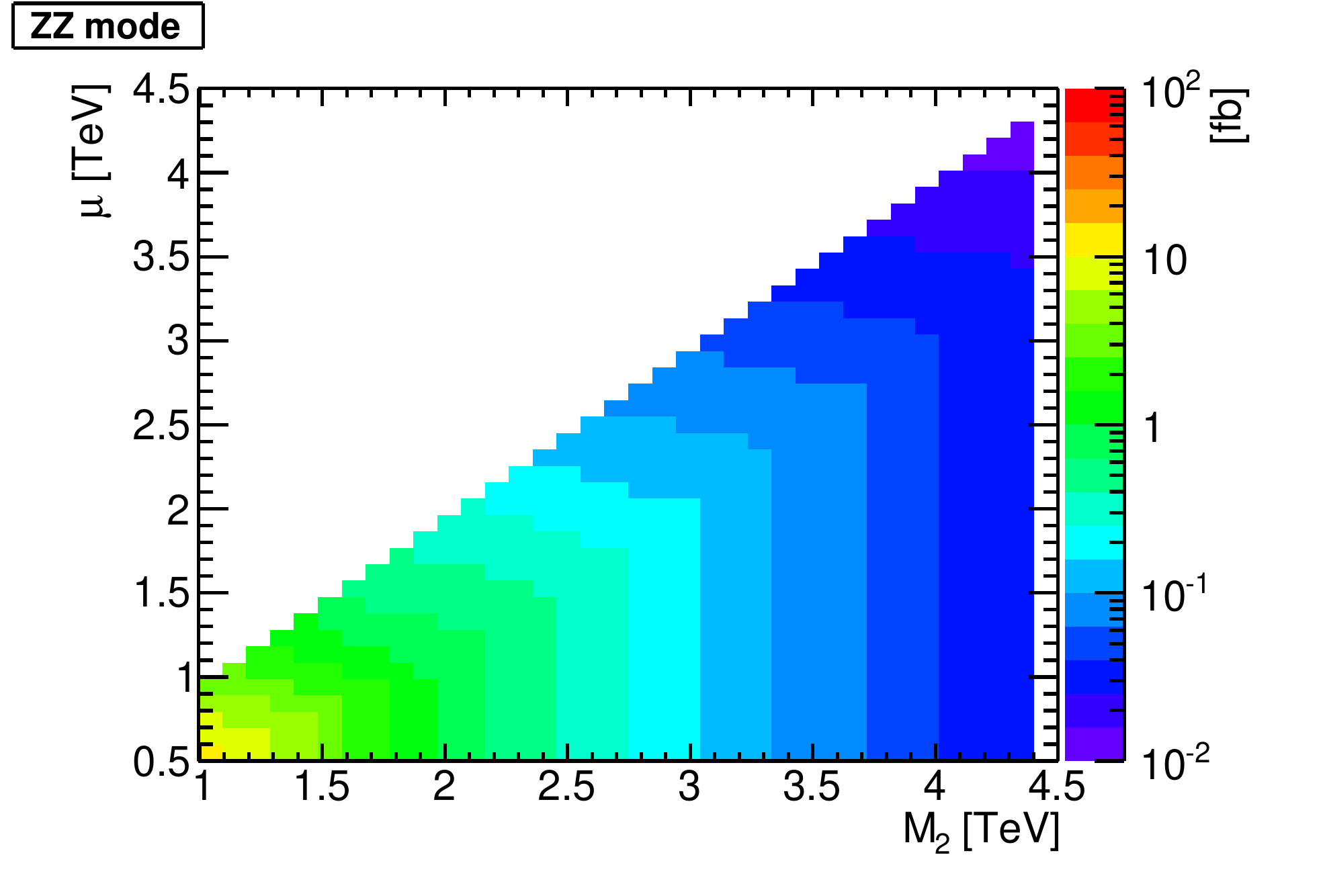}
		\includegraphics[width=0.45\textwidth]{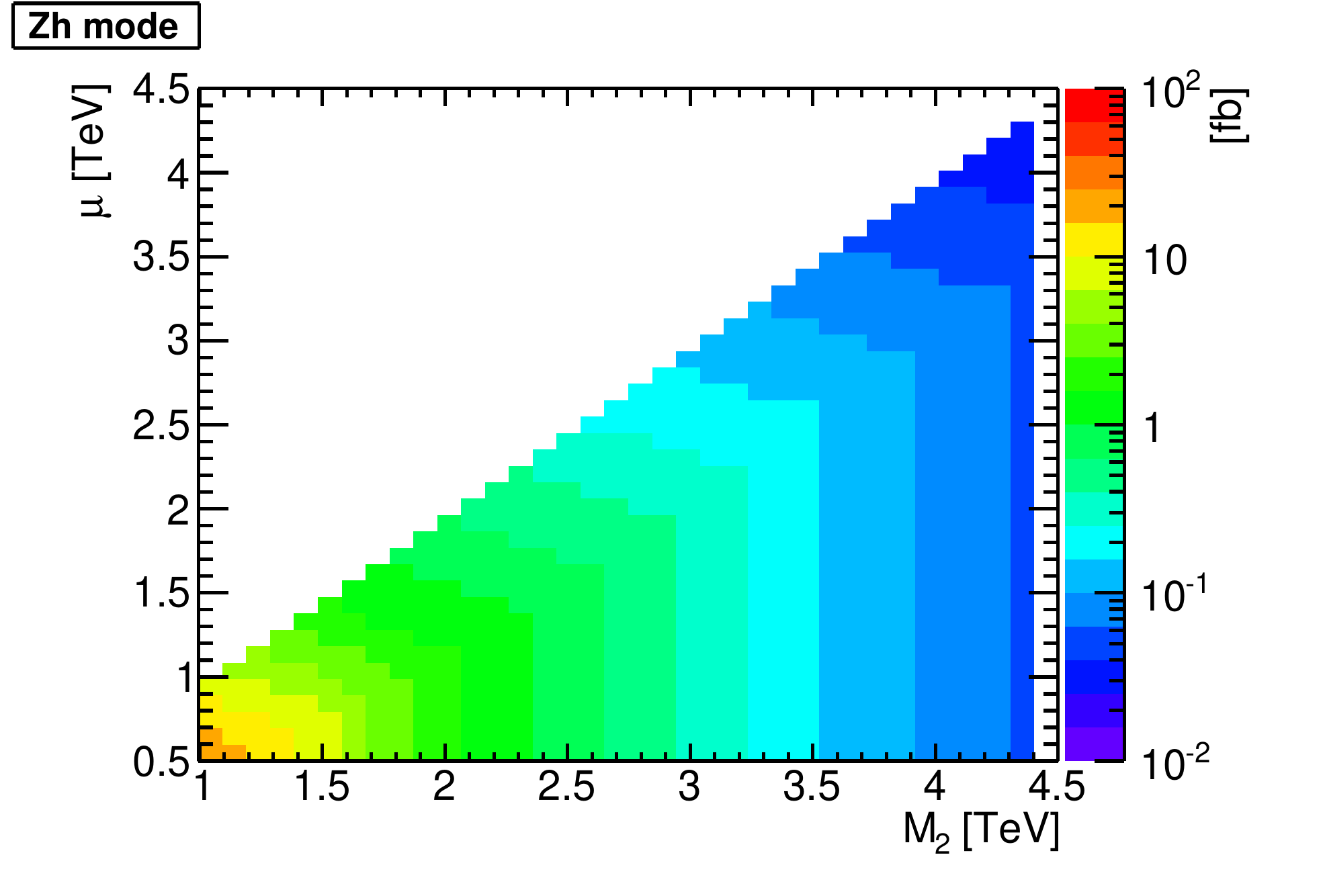}
		\includegraphics[width=0.45\textwidth]{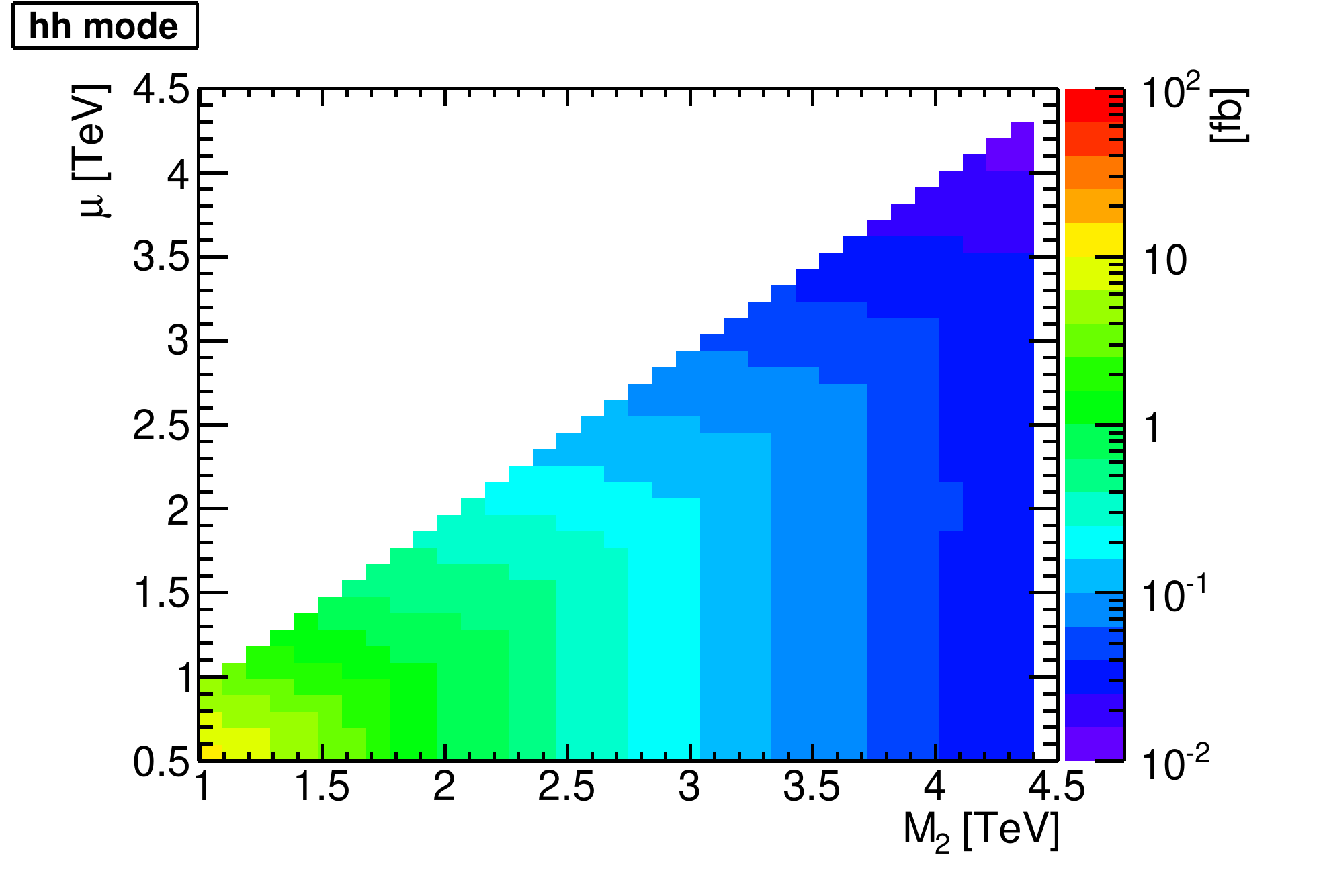}				
	\caption{ 
	The cross sections of the 6 distinguishable modes,  $\chi^\prime \chi^\prime \to XY \chi \chi$ with $XY = WZ, Wh, WW, ZZ, Zh$ and $hh$,
	as functions of $M_2$ and $\mu$.  SUSY particles other than $W$-inos and Higgsinos are decoupled.  
\label{fig:modes}}
\end{figure}

Fig.~\ref{fig:br} shows the branching ratios of $\tilW^\pm$ and $\tilW^0$, which have been calculated using {\tt SUSY-HIT} \cite{Djouadi:2006bz}.  
One can see that the branching ratios approach Eq.~(\ref{eq:br}) in the large $M_2$ limit.  
For the region where $|M_2 - \mu|$ is close to the masses of SM bosons, the decay mode into $W^\pm$ enhances since it has
the largest phase space factor.

Since the charged and neutral $W$-inos are almost mass degenerate, it may not be possible to resolve $\tilW^\pm \to X Y$ and $\tilW^0 \to X^\prime Y^\prime$ in hadron colliders if $XY$ is equal to $X^\prime Y^\prime$ up to soft activities.
Similarly, four degenerate Higgsinos would not be resolvable, since $\tilH^\pm$ and $\tilH_2^0$ usually decay promptly into $\tilH_1^0$ and their decay products are too soft to be detected.     
We therefore categorise the processes into distinguishable groups in terms of the SM bosons appearing in the final states.
For example, $\chi^\prime \chi^\prime \to W Z \chi \chi$ process ($WZ$ mode) includes
$\tilW^+ \tilW^- \to (W^\pm \tilH_{1/2}^0) (Z \tilH^\mp)$, $\tilW^\pm \tilW^0 \to (W^\pm \tilH_{1/2}^0) (Z \tilH_{1/2}^0), \, (Z \tilH^\pm) (W^\pm \tilH^\mp)$ 
and $\tilW^0 \tilW^0 \to (W^\pm \tilH^\mp)(Z \tilH_{1/2}^0)$.
We show the cross sections of the all 6 distinguishable modes, $WZ$, $Wh$, $WW$, $ZZ$, $Zh$ and $hh$ modes, in the $M_2 - \mu$ plane in Fig.~\ref{fig:modes}.

\begin{figure}[t!]
\begin{center}
 \includegraphics[width=0.55\textwidth]{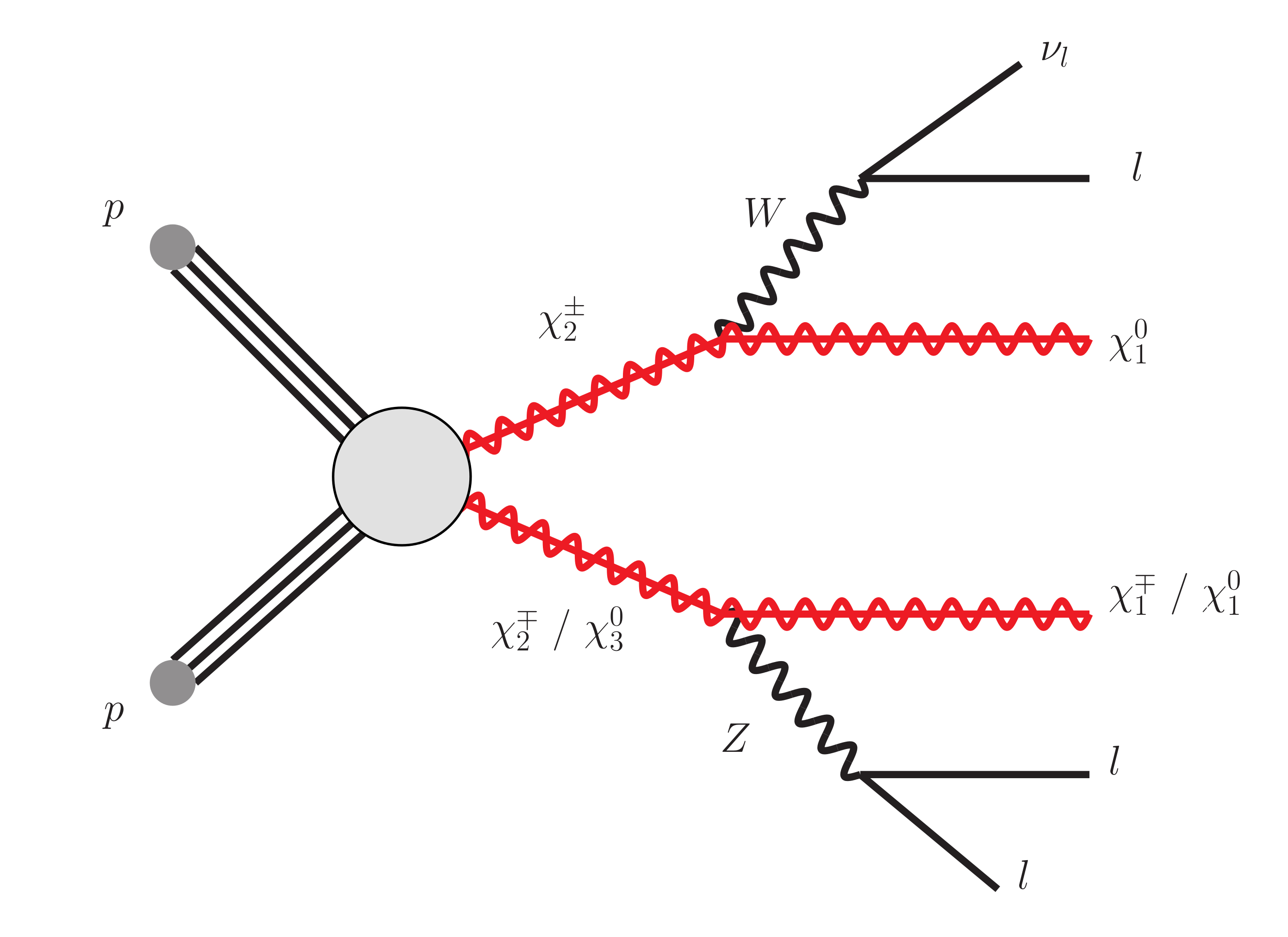}
\caption{The dominant event topology for signal events.}\label{fig:Feyn-WZ}
\end{center}
\end{figure}

One can see that the modes containing at least one $W$ have considerably larger cross sections compared to the others at the same mass point.  
In particular, the $WZ$ mode is promising\footnote{
The $Wh$ mode is also interesting.  See \cite{Baer:2012ts, Han:2013kza, Ghosh:2012mc, Byakti:2012qk, Papaefstathiou:2014oja} for some recent studies.
} 
because one can reduce the QCD and $t \bar t$ backgrounds significantly by requiring three high $p_T$ leptons (See Fig.~\ref{fig:Feyn-WZ}.).
%Hereafter, we concentrate our study on the three lepton channel in the $WZ$ mode.
%
Taking advantage of this we henceforth study the expected discovery reach and exclusion limit for chargino-neutralino production in the $WZ$ mode.

\begin{figure}[t!]
\begin{center}
 \includegraphics[width=0.6\textwidth]{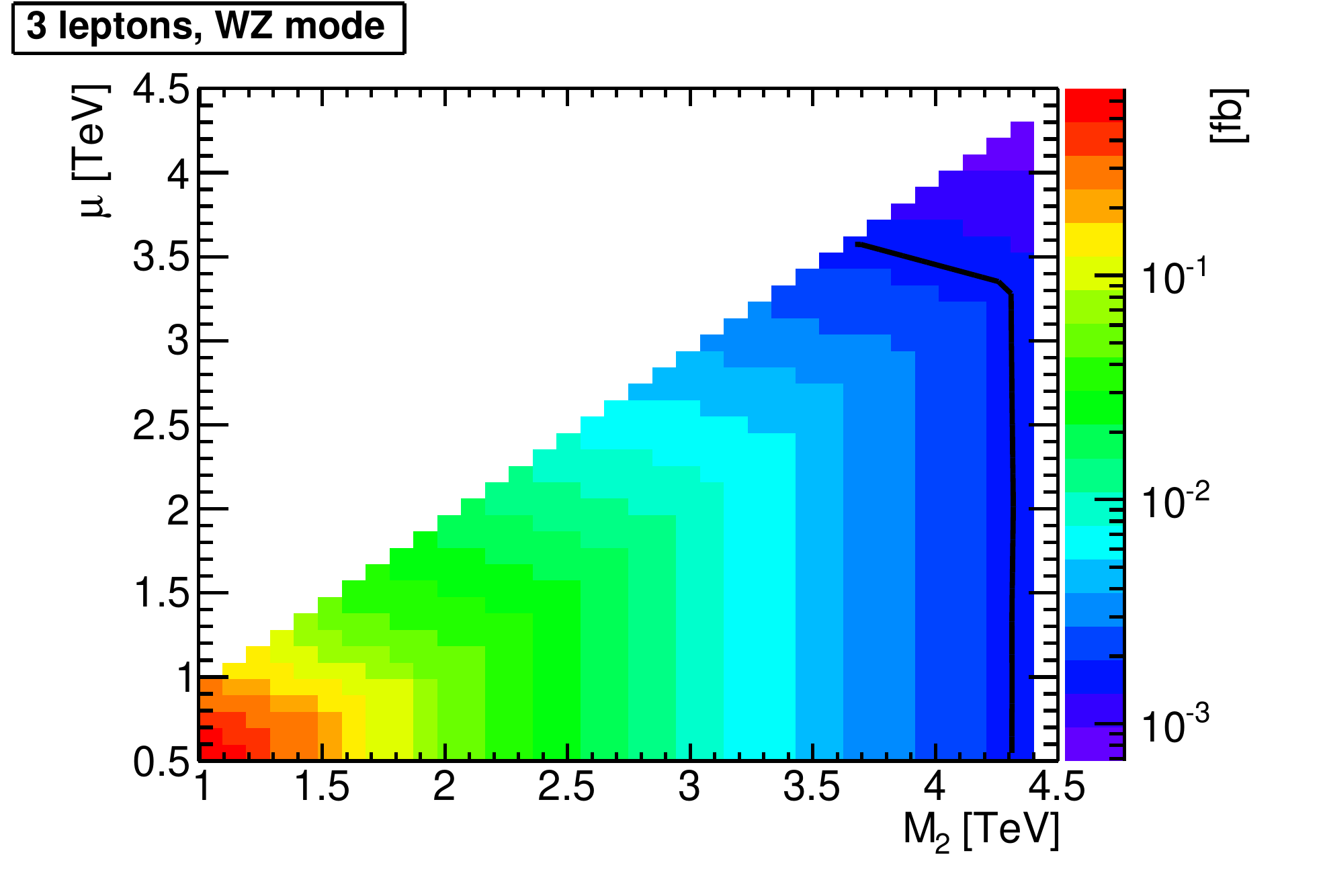}
\caption{The cross section of $\chi^\prime \chi^\prime \to WZ \chi \chi \to 3\, \ell \nu \chi \chi$ as a function of $M_2$ and $\mu$. The black curve represents the limit beyond which less than 5 signal events are produced, assuming the integrated luminosity of 3000 fb$^{-1}$. %SUSY particles other than $W$-inos and Higgsinos are decoupled.}
\label{fig:WZ-lept}
}
\end{center}
\end{figure}

In Fig.~\ref{fig:WZ-lept}, we show the cross section of the $WZ$ mode after taking account of the branching ratios of the gauge bosons into $3 \ell + \nu$.
The black curve represents the limit beyond which less than 5 signal events ($\chi^\prime \chi^\prime \to WZ \chi \chi \to 3 \ell \nu \chi \chi$) are produced, assuming the integrated luminosity of 3000 fb$^{-1}$.
This provides 
a rough estimate of the theoretically maximum possible exclusion limit 
assuming zero background with perfect signal efficiency.
%The black curve corresponds to the mass points where the expected signal yield ($\chi^\prime \chi^\prime \to WZ \chi \chi \to 3 \ell \nu \chi \chi$) is 5 assuming 3000 fb$^{-1}$ of integrated luminosity, which provides a rough estimate of the theoretically maximum possible exclusion limit 
%assuming zero background with perfect signal efficiency.
\color{black}

%Our study will focus on the three lepton final state. Before describing our analysis, it is useful to consider the maximum possible discovery reach assuming 100\% reconstruction efficiency and a background of order one event. To this end, Fig.~\ref{fig:WZ-lept} shows the cross section times branching ratio for the fully leptonic $WZ$ decays.
%The black curve is the line of ${5 \over 3000}$~fb, i.e. corresponds to five signal events.

%
\section{The simulation setup \label{sec:sim}}
  
We use the {\sf Snowmass} background samples \cite{Anderson:2013kxz} to estimate the Standard Model (SM) backgrounds.
We include the relevant SM processes, which are summarised in Table~\ref{tab:SMBG}.   
%The most relevant SM processes that contributes are: 1) diboson production, especially direct $WZ$ production. We take {\sf Snowmass} diboson samples VV, to include this. 2) top-pair + gauge boson production, included in ttV {\sf Snowmass} sample. 3) triple boson production, included in VVV {\sf Snowmass} sample. 4) top + gauge boson production, included in tV {\sf Snowmass} sample. 
%For background events, we use SM background samples provided by Snowmass \cite{Anderson:2013kxz}. 
%These samples were made using customise {\tt Delphes\,3} \cite{deFavereau:2013fsa}. 

For signal events we first generate chargino and neutralino production events using {\tt MadGraph\,5} with the parameters obtained by {\tt SUSY-HIT}. We consider two production processes $pp \to \chi^{+}_{2}\chi^{-}_{2}$ and $pp \to \chi^{\pm}_{2}\chi^{0}_{3}$, where $\chi^\pm_{2} \sim \widetilde W^\pm$ and $\chi^{0}_{3} \sim \widetilde W^0$. 
%The pair production of $\chi^{0}_{3}\chi^{0}_{3}$ is irrelevant in the parameter space we consider. 
The generated samples are then passed to {\tt BRIDGE} \cite{Meade:2007js} to have the charginos and neutralinos decay. 
%according to the branching ratios calculated by {\tt SUSY-HIT}. 
We then only accept the events with $W$ and $Z$ in the final states, and pass those events once again to {\tt BRIDGE} to let $W$ and $Z$ decay leptonicaly.
Finally we simulate the effects of parton shower, hadronization and detector resolutions using {\tt Pythia\,6} \cite{Sjostrand:2006za} and {\tt Delphes\,3} \cite{deFavereau:2013fsa}.
The detector simulation is tuned according to the {\sf Snowmass} detector framework \cite{Anderson:2013kxz}. 

\begin{table}[t!]
\begin{center}
\begin{tabular}{|c|c|c|c|}
\hline
Name & Snowmass & Relavant sub-processes & $\sigma_{\rm total}^{\rm NLO}$ [pb] \\ 
\hline
\multirow{1}{*}{diboson} & \multirow{1}{*}{VV} & $W^+ W^-$, $W^\pm Z$, $ZZ$ & \multirow{1}{*}{430.5} \\
\multirow{1}{*}{top-pair + gauge boson} & \multirow{1}{*}{ttV} & \multirow{1}{*}{$t \bar t W^\pm$, $t \bar t Z$, $t \bar t\, h$} & \multirow{1}{*}{219.9} \\
\multirow{1}{*}{top + gauge boson} & \multirow{1}{*}{tV} & \multirow{1}{*}{$t W^\pm$,  $\bar t W^\pm$} & \multirow{1}{*}{182.5} \\
\multirow{1}{*}{triple gauge boson} & \multirow{1}{*}{VVV} & $W^+ W^- W^\pm$, $W^+ W^- Z$, $W^\pm ZZ$, $ZZZ$ & \multirow{1}{*}{36.4} \\
\hline
\end{tabular}
\end{center}
\caption{The Standard Model background included in the analysis. For each background category, we only list sub-processes relevant in the 3 lepton analysis. Reported cross sections include all sub-processes in corresponding background categories. 
\label{tab:SMBG}
}
\end{table}%

\section{The kinematic distributions \label{sec:distributions}}

\begin{figure}[t]
\begin{center}
  \subfigure[]{\includegraphics[width=0.48\textwidth]{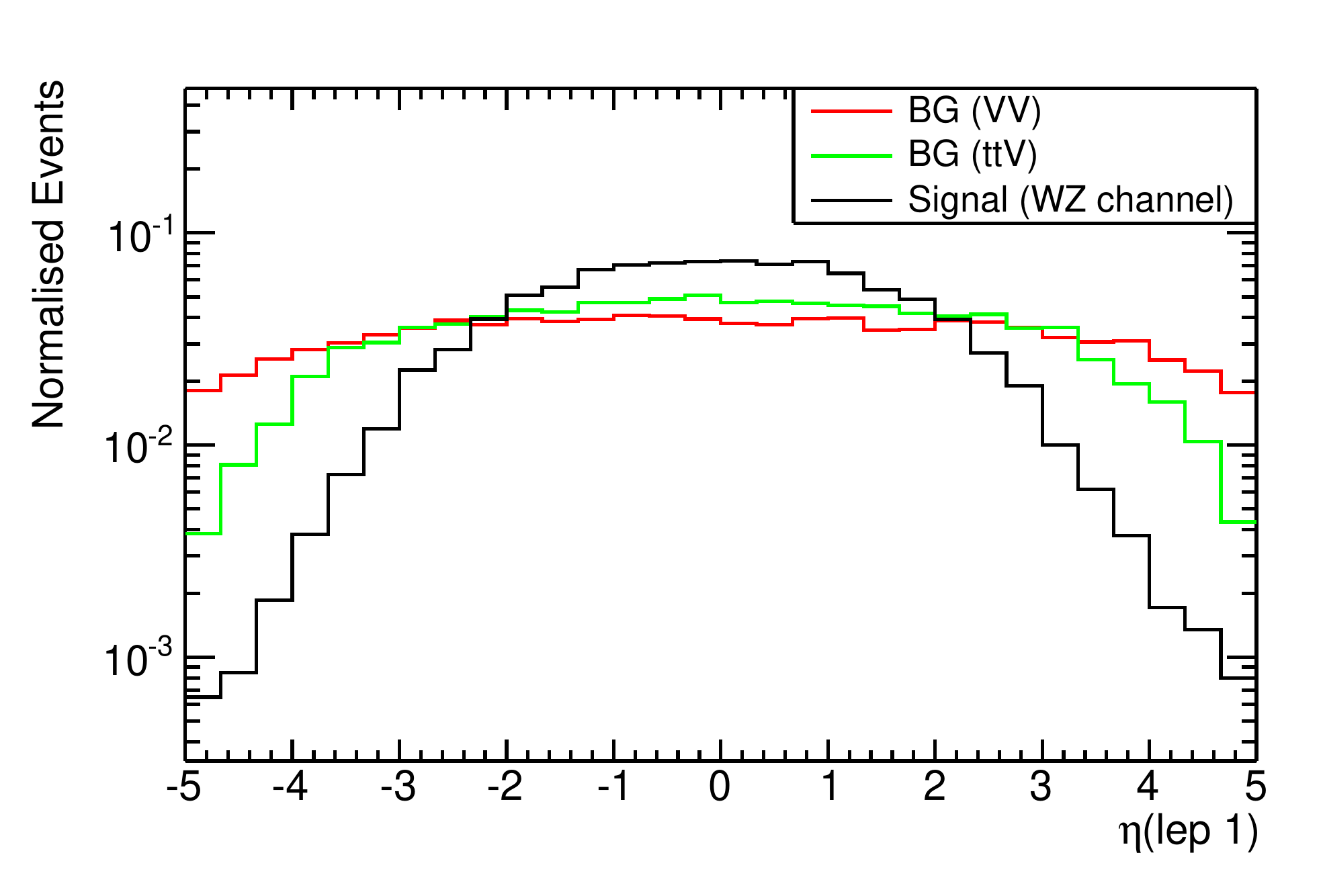}\label{fig:eta}} 
  \subfigure[]{\includegraphics[width=0.48\textwidth]{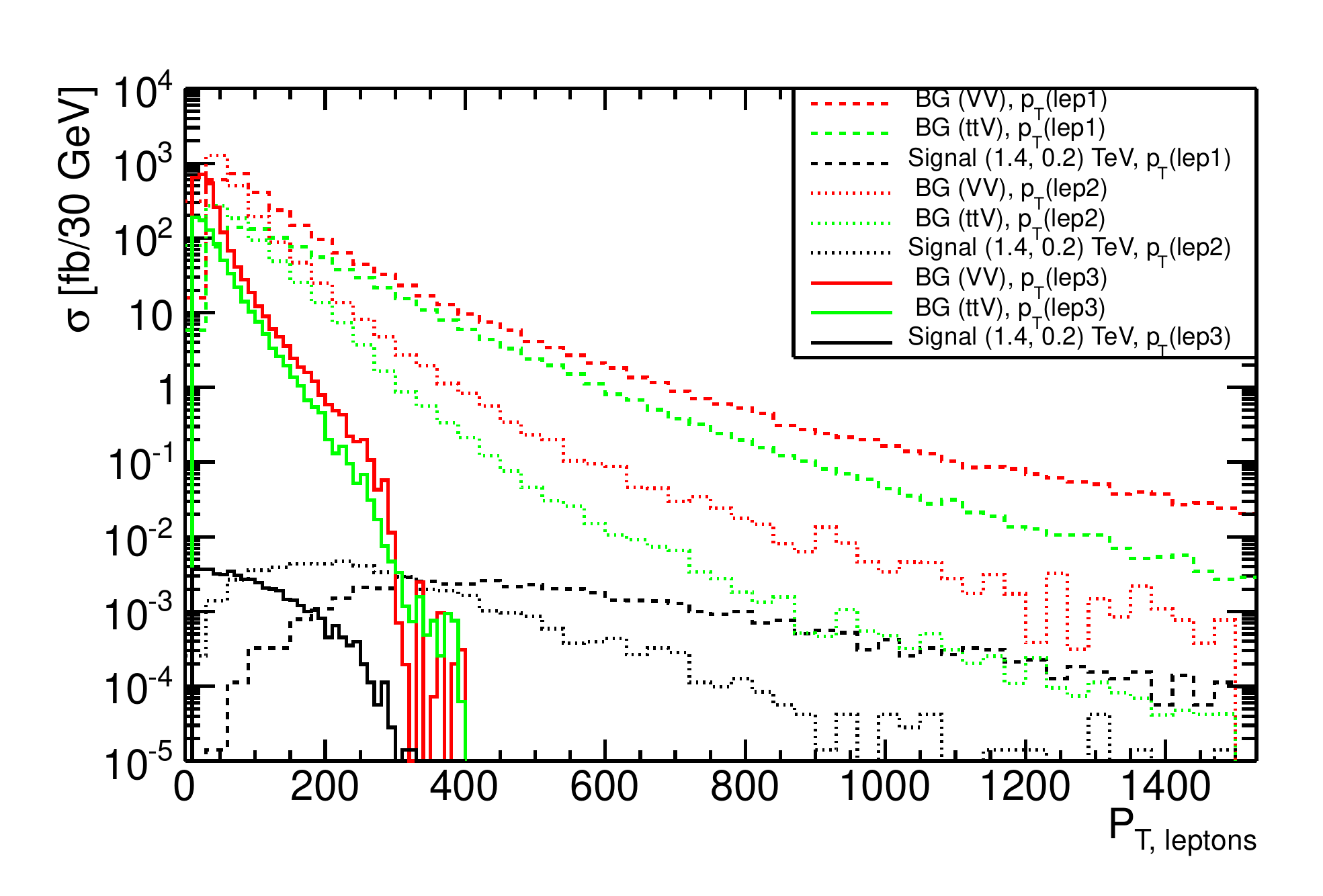}\label{fig:pT}} \hspace{0.2cm}
  \subfigure[]{\includegraphics[width=0.48\textwidth]{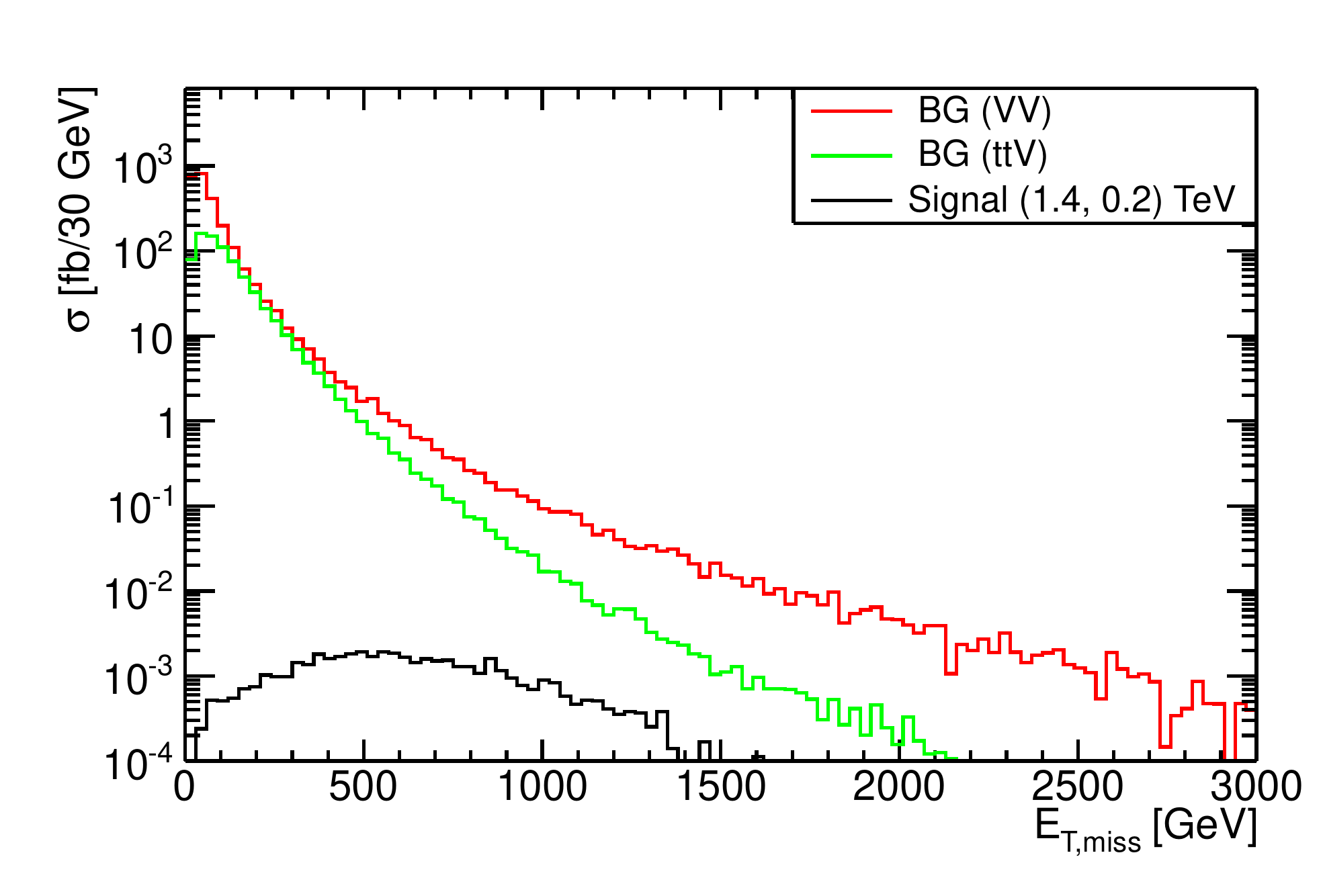}\label{fig:met}} 
  \subfigure[]{\includegraphics[width=0.48\textwidth]{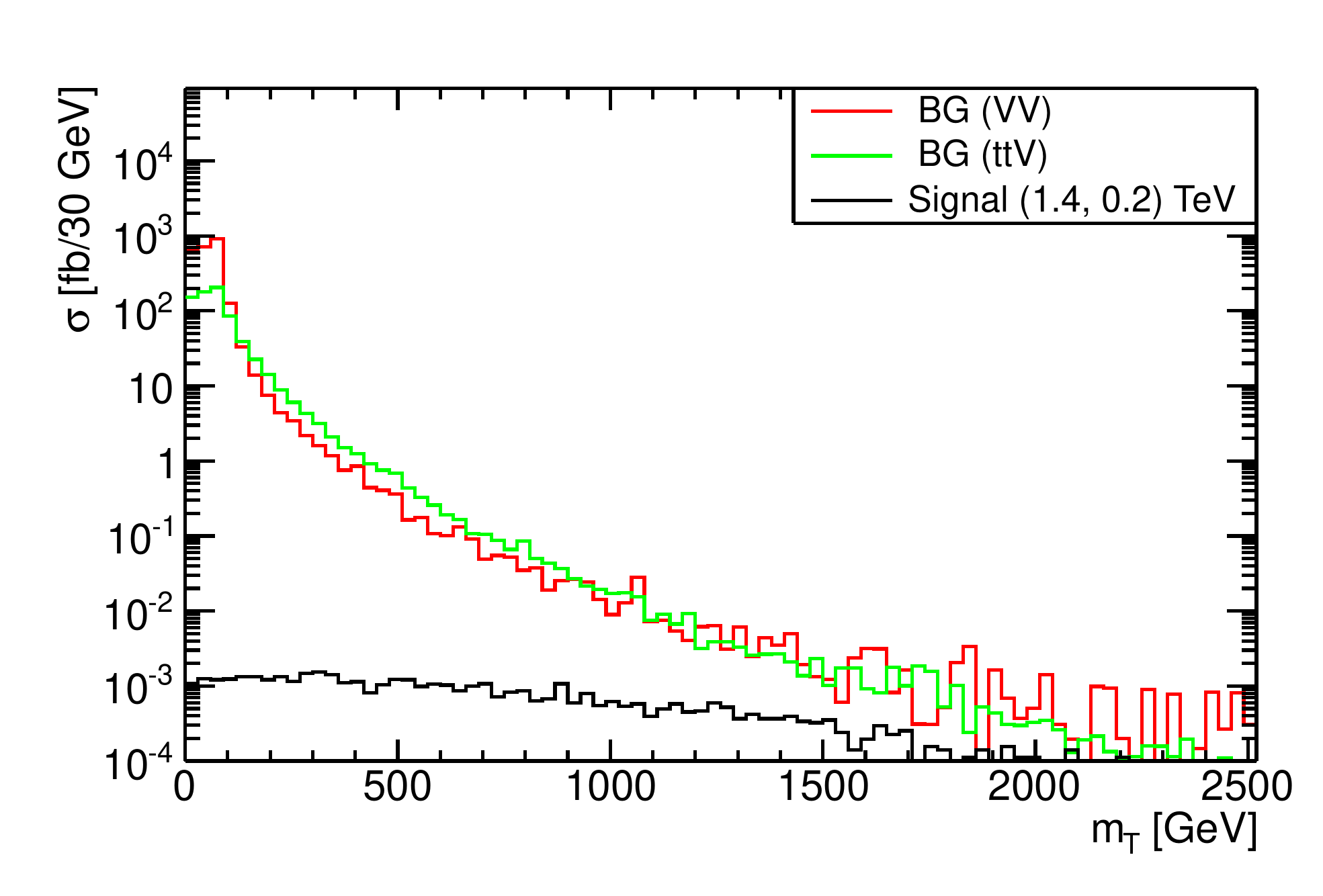}\label{fig:mT}} %\hspace{0.2cm}
\caption{  The distributions of {\bf (a)} the leading lepton pseudo-rapidity, $\eta_{\ell_1}$, {\bf (b)} $p_T$ of the three hardest leptons, {\bf (c)} the missing transverse energy, $\met$,
{\bf (d)} the transverse mass, $m_T$. 
The backgrounds are diboson (VV) and associated top-pair plus vector boson production (ttV). 
The signal events are generated at our benchmark point, $M_2 = 1.4$ TeV and $\mu = 200$ GeV, and
only $WZ$ mode is considered.
The parton level events are used for {\bf (a)}, whilst 
the detector level events after applying the 3 lepton + SFOS cuts are used for {\bf (b)}, {\bf (c)} and {\bf (d)}.
%The black histograms and dots corresponds to the $\chi^\prime \chi^\prime \to W^\pm Z \chi \chi$ signal with $M_2 = 1.4$ TeV, $\mu = 200$ GeV. 
%The red ones are for the $WZ$ background.
\label{fig:distributions} }
\end{center}
\end{figure}

In this section we show some kinematic distributions for the background and signal events. 
We consider the $WZ$ mode for signal and 
diboson (VV) and top-pair plus gauge boson (ttV) processes for backgrounds.
The signal distributions are generated at a benchmark point: $M_2 = 1.4$ TeV, $\mu = 200$ GeV.  
%The details of our computer simulation methods are described in Section \ref{sec:limit}.
Throughout this section we use a notation denoting the $i$-th hardest lepton (electron or muon) by $\ell_i$ (namely, $p_T(\ell_i) > p_T(\ell_j)$ for $i < j$).

Fig.~\ref{fig:eta} shows the normalised distributions of the leading lepton pseudo-rapidity, $\eta_{\ell_1}$, for signal (black) and background (red for VV and green for ttV).
The distributions are obtained at a parton level without selection cuts apart from $p_T(\ell_1) > 10$ GeV
to understand the bare distribution before taking the detector acceptance into account.  
One can see that the leptons in the background tend to be more forward compared to the signal leptons.
The production threshold is much lower for the backgrounds and more asymmetric momentum configurations are allowed for the initial partons.
If one of the initial partons has a much larger momentum than the other, the system is boosted in the direction of the beam pipe and the
leptons tend to be produced in the forward region.\footnote{
For the $W^+Z$ background, the initial state is often $u$ and $\bar d$.  
If the partonic collision energy is much smaller than the proton-proton collision energy, it is more likely to find a valence quark $u$ carrying a larger fraction of the proton momentum 
compared to the sea quark $\bar d$.
}  
Another effect is as follows.  
Unlike the signal, production of the backgrounds have a contribution from $t$-channel diagrams. 
In 100 TeV colliders, the SM gauge bosons can effectively be regarded as ``massless" particles and
there is an enhancement in the region of the phase space where the gauge bosons are produced in the forward region. 

Fig.~\ref{fig:pT} shows the $p_T$ distributions of the three hardest leptons.
The distributions are obtained after taking the hadronization and detector effects into account
and requiring at least 3 leptons (with $p_T > 10$ GeV, $|\eta| < 2.5$), of which two are same flavour and opposite sign (SFOS).   
As can be seen, the $p_T$-spectrum of background leptons has peaks below 100 GeV, whilst 
the signal peaks at around 300, 150 and $\lsim 50$ GeV for the leading, second leading and third leading leptons for our benchmark point.  

We also show the $\met$ distributions in Fig.~\ref{fig:met}, where we use the same event sample as those in Fig.~\ref{fig:pT}.
The main source of the $\met$ in the background are the neutrinos produced from $W$ and $Z$ decays and 
the distribution has a peak around $30 - 40$ GeV.
Above this peak, the background $\met$ distribution falls quickly.
On the other hand, a large $\met$ can be produced from the signal from the decays of heavy charginos and neutralinos.
The typical scale of $\met$ is given by $\sim M_2/2$.
As can be seen, the signal distribution has a peak around 500 GeV.
This indicates that a hard cut on $\met$ will greatly help to improve the signal to background ratio.
   
We show the transverse mass $m_T$ distributions in Fig.~\ref{fig:mT}, where the event samples are again the same as those used in Fig.~\ref{fig:pT}.
We define $m_T \equiv \sqrt{2 |p_T(\ell^\prime)||\met|(1 - \cos\Delta\phi)}$, where $\ell^\prime$ is the hardest lepton amongst those 
not chosen as the SFOS lepton pair and $\Delta \phi$ is the azimuthal difference between the $\ell^\prime$ and the direction of $\overrightarrow{p}_T^{\rm miss}$. %$\met$.     
In the $WZ$ background, this distribution has an endpoint at $m_W$
and above the endpoint the distribution drops very sharply.
In the signal events, the distributions are much broader, as can be seen in Fig.~\ref{fig:mT}.
A harsh cut on $m_T$ would also be very helpful to reject a large fraction of background without sacrificing too many signal events.

\section{The limit and discovery reach \label{sec:limit}}

\subsection{The event selection \label{sec:selection}}

Our event selection consists of two parts: $preselection$ and $signal~region~(SR)~selection$.  
The $preselection$ requirement is:
\begin{itemize}
\item exactly three isolated leptons with $p_T > 10$~GeV and $|\eta| < 2.5$
\item a same-flavour opposite-sign (SFOS) lepton pair with $|m_{\ell \ell}^{\rm SFOS} - m_Z| < 10$ GeV
\item no $b$-tagged jet
\end{itemize}
With the first condition one can effectively reject the QCD, hadronic $t \bar t$ and single gauge boson backgrounds.
The definition of lepton isolation and some discussion around it is given in Appendix~\ref{sec:isolation}.
The second condition is introduced to remove the leptonic SM processes without $Z$ bosons, such as $t \bar t W^\pm$ and $W^+ W^- W^\pm$.
The last condition is effective to reduce the SM backgrounds containing top quarks.
In the simulation we use the $b$-tagging efficiency of about $70\, \%$, which is set in the {\tt Delphes} card used in the {\sf Snowmass} backgrounds.  

\begin{table}[t!]
\begin{center}
\begin{tabular}{c|c|c|c}
\hline 
Signal Region & 3 lepton $p_T$ [GeV] & $E^{\mathrm{miss}}_T$ [GeV] & $m_{T}$ [GeV] \\ 
\hline
Loose  &  $> 100, 50, 10$  &  $> 150$  &  $> 150$ \\
Medium &  $ > 250, 150, 50 $  &  $> 350$  &  $> 300$ \\
Tight  &  $ > 400, 200, 75 $  &  $> 800$  &  $> 1100$ \\
\hline
\end{tabular}
\end{center}
\caption{The event selection cuts required in the signal regions. These cuts are applied on top of the preselection cuts.}
\label{tab:SRs}
\end{table}%

In order to obtain as large coverage as possible in the $M_2 - \mu$ parameter plane, we define three signal regions: {\it Loose}, {\it Medium}, {\it Tight}.
These signal regions are defined in Table~\ref{tab:SRs}.
The selection cuts are inspired by the kinematical distributions shown in Fig.~\ref{fig:distributions}. The
{\it Loose} region, which has the mildest cuts, is designed to constrain the degenerate mass region ($M_2 \gsim \mu$), whereas the
{\it Tight} region, which has the hardest cuts, targets the hierarchical mass region ($M_2 \gg \mu$). The
{\it Medium} region is also necessary to extend the coverage in the intermediate mass region.

The visible cross section (the cross section for the events satisfying the event selection requirements) for each signal region 
is shown in Appendix~\ref{sec:cutflow}.
The information for the detailed breakdown of the background contribution and the visible cross section at each step of the selection is also shown. 
The number of total background events are expected to be 38400, 810 and 12.3 for the {\it Loose}, {\it Medium} and ${\it Tight}$ signal regions, respectively,
at 3000 fb$^{-1}$ of integrated luminosity.

\subsection{The result \label{sec:result}}

\begin{figure}[t]
\begin{center}
  \subfigure[]{\includegraphics[width=0.48\textwidth]{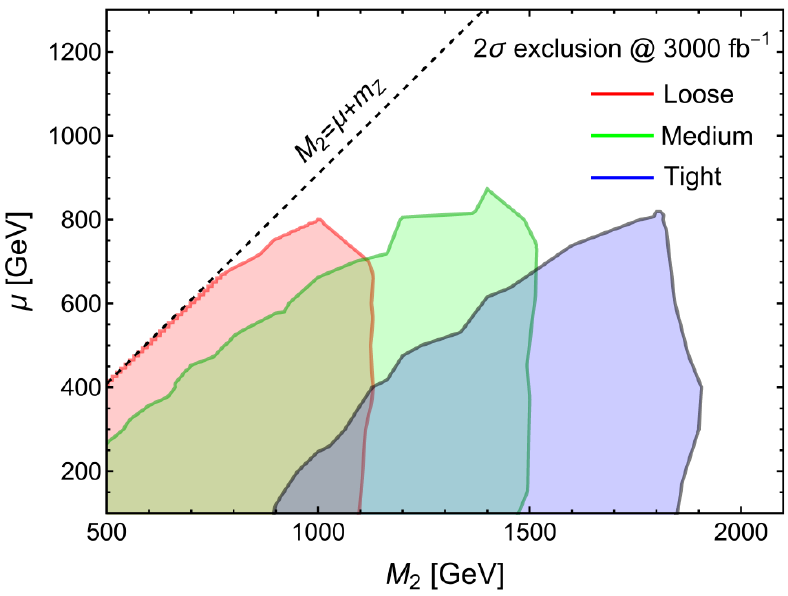}\label{fig:excl3000}} \hspace{0.2cm}
  \subfigure[]{\includegraphics[width=0.48\textwidth]{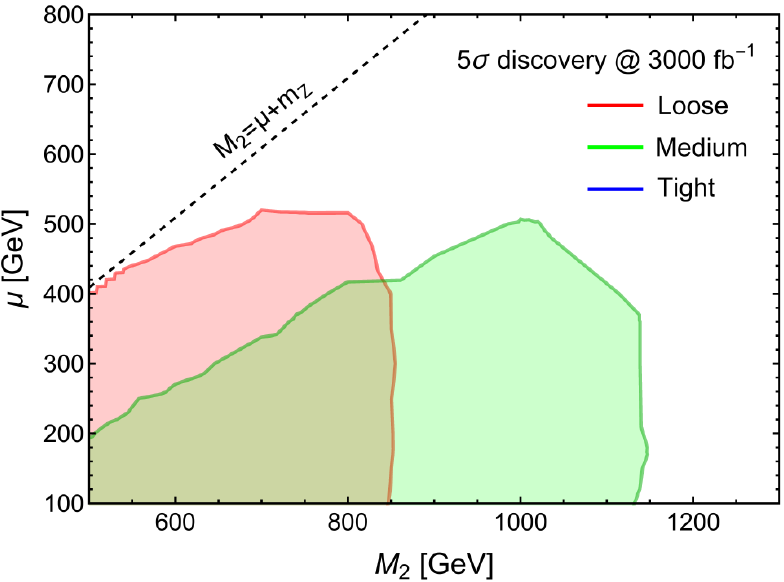}\label{fig:disc3000}} \hspace{0.2cm}
\caption{The exclusion limits {(\bf a)} and the discovery reaches {(\bf b)} obtained from three signal regions.  The integrated luminosity of 3000 fb$^{-1}$ is assumed.}   
 \label{fig:reach3000}
\end{center}
\end{figure}

\begin{figure}[t]
\begin{center}
  \subfigure[]{\includegraphics[width=0.48\textwidth]{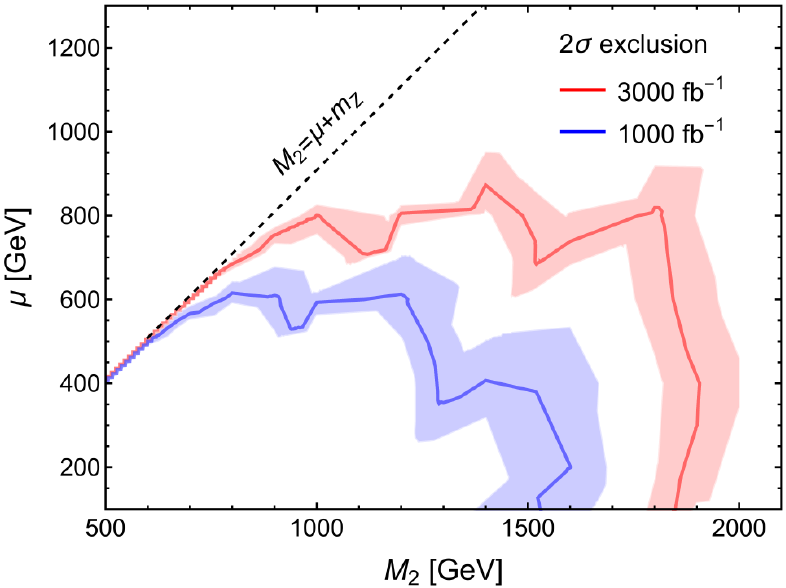}\label{fig:excl}} \hspace{0.2cm}
  \subfigure[]{\includegraphics[width=0.48\textwidth]{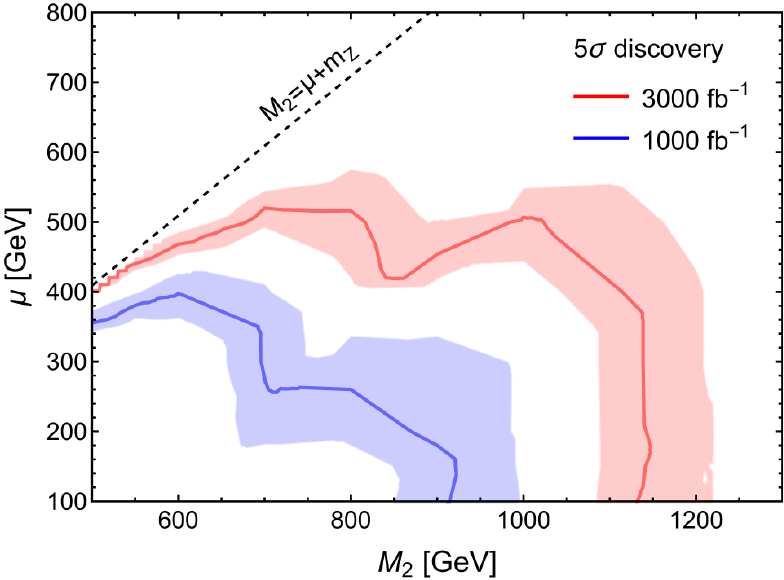}\label{fig:disc}} \hspace{0.2cm}
\caption{The global exclusion limits {\bf (a)} and the discovery reaches {\bf (b)} for 3000 fb$^{-1}$ (red) and 1000 fb$^{-1}$ (blue).
The shaded region represent the uncertainty when varying the background yield by $30 \, \%$.
}    
 \label{fig:reach}
\end{center}
\end{figure}

In Fig.~\ref{fig:excl3000}, we show the $2\, \sigma$ exclusion limits in the $\mu - M_2$ parameter plane obtained by the different signal regions.
The shaded regions have $S/\sqrt{B} \geq 2$, where $S$ and $B$ are the number of expected signal and background events falling into the signal regions, respectively.
For signal we use a constant $k$-factor of 1.3 across the parameter plane.
One can see that the three signal regions are complementary and $M_2$ can be constrained up to $\sim 1.8$ TeV for $\mu \lsim 800$ GeV.  

Fig.~\ref{fig:disc3000} shows the $5 \, \sigma$ discovery reach ($S/\sqrt{B} \geq 5$) obtained from the different signal regions.
As can be seen, the {\it Loose} and {\it Medium} signal regions provide the discovery reach up to about $850$ and $1.1$ TeV, respectively, for $\mu \lsim 450$ GeV.
On the other hand, the {\it Tight} signal region does not have sensitivity to $S/\sqrt{B} \geq 5$.

We show in Fig.~\ref{fig:excl} the global $2 \, \sigma$ exclusion limits for integrated luminosities of 3000 fb$^{-1}$ (red) and 1000 fb$^{-1}$ (blue). 
The global exclusion limit is obtained by choosing the signal region that provides the largest $S/\sqrt{B}$ for each mass point.
The shaded regions around the solid curves represent the uncertainty when varying the background yields by $\pm 30\,\%$. 
One can see that changing the background by $30\,\%$ results in a $\sim 100$ GeV shift in $M_2$ for the $\mu \ll M_2$ region.
$M_2$ can be constrained up to 1.8 TeV with $\mu \lsim 800$ GeV for 3000 fb$^{-1}$, which can be compared with 
the projected chargino neutralino mass limit of 1.1 TeV for the high luminosity LHC with 3000 fb$^{-1}$ obtained by ATLAS \cite{ATLAS:2014projection}.
For 1000 fb$^{-1}$ the limit on $M_2$ is about 1.5 TeV with $\mu \lsim 400$ GeV as can be seen in Fig.~\ref{fig:excl}.

Fig.~\ref{fig:disc} shows the global $5 \, \sigma$ discovery reach for 3000 fb$^{-1}$ (red) and 1000 fb$^{-1}$ (blue) with the 30$\, \%$ uncertainty bands for background.
One can see that charginos and neutralinos can be discovered up to $M_2 \lsim 1.1$ TeV with $\mu \lsim 500$ GeV for 3000 fb$^{-1}$ integrated luminosity, which can be compared with the projected ATLAS value of 0.8 TeV for the high luminosity LHC \cite{ATLAS:2014projection}.
For 1000 fb$^{-1}$, charginos and neutralinos can be discovered up to  900 GeV with $\mu \lsim 250$ GeV.

\section{Conclusion \label{sec:concl}}

In this paper we studied the prospect of chargino and neutralino searches at a 100 TeV $pp$ collider assuming 3000 (1000) fb$^{-1}$ of integrated luminosity.
Our particular focus was the case where the Higgsinos form the lightest SUSY states (the lightest charginos and the two lightest neutralinos, which are almost mass degenerate) and $W$-inos form the second lightest states (the heavier charginos and the third lightest neutralino, which are almost mass degenerate). 
The other SUSY particles including $B$-ino are assumed to be decoupled, which is partly motivated by the current LHC results as well as popular scenarios of SUSY breaking and its mediation.  We have shown that in this situation the LO production cross sections of 2 TeV $W$-inos are as large as 100 fb$^{-1}$ and the branching ratio of $W$-inos follows a simple formula, which can be derived from the Goldstone equivalence theorem.  

From a study of kinematic distributions of signal and background we found harsh cuts on lepton $p_T$ ($> 50 - 400$ GeV), $E_T^{\rm miss}$ ($> 150 - 800$ GeV) and
$m_T$ ($> 150 - 1100$ GeV) are beneficial to improve the signal and background ratio and designed three complementary signal regions.
Using these three signal regions, we found the $5\, \sigma$ discovery reach ($2\,\sigma$ exclusion limit) for the chargino-neutralino mass is 1.1 (1.8) TeV for $\mu \lsim 500$ (800) GeV, which can be compared with the projected LHC reach (limit) of 0.8 (1.1) TeV obtained by ATLAS \cite{ATLAS:2014projection}.   
For 1000 fb$^{-1}$ the discovery reach (exclusion limit) for the chargino-neutralino mass is found to be 0.9 (1.5) TeV for $\mu \lsim 250$ (400) GeV.

\acknowledgments
We would like to thank Lian-Tao Wang, Stefania Gori and Sunghoon Jung for discussions.
The work of BSA is supported by the UK STFC via the research grant ST/J002798/1. BSA also acknowledges the hospitality of the Center for the Future for High Energy Physics, Beijing, where this work was first presented.
The work of K.S. was supported in part by
the London Centre for Terauniverse Studies (LCTS), using funding from
the European Research Council 
via the Advanced Investigator Grant 267352. The work of KB is supported by a Graduate Teaching Assistantship from King's College London.

\newpage

\appendix

\section{The lepton isolation requirement}\label{sec:isolation}

In hadron colliders, leptons (electrons and muons) may arise from heavy hadron decays.    
Those ``background'' leptons are usually found together with other particles around them.
The leptons originating from gauge boson decays can therefore be distinguished from the background leptons 
by investigating activity around the lepton. 
For this check, {\tt Delphes 3} uses an isolation variable, $I$, defined as
\beq
I( \ell )=\frac{\sum\limits_{i\neq \ell}^{\Delta R<R,\ p_{T}(i)>p^{\rm min}_{T}} p_{T}(i)}{p_{T}( \ell )},
\eeq
where the numerator sums the $p_T$ of all  particles (except for the lepton itself) with $p_T > p_T^{\rm min}$ lying within a cone of radius $R$ around the lepton.
If $I(\ell)$ is smaller than $I_{\rm min}$, the lepton is said to be isolated, otherwise gets rejected as background.  
The {\sf Snowmass} samples were generated using {\tt Delphes 3} with the lepton isolation parameters of $R=0.3$, $p^{\rm min}_{T}=0.5$ and $I_{min}=0.1$.

A 100 TeV collider can explore charginos and neutralinos with their mass scale of a few TeV.
If the mass hierarchy between $W$-ino states and Higgsino states are much higher than the gauge bosons mass scale,
the $W$ and $Z$ produced from the $W$-ino decays will be highly boosted.
If such a boosted $Z$ decays into a pair of same-flavour opposite-sign (SFOS) leptons, those two leptons can be highly collimated, and one may be
rejected by the isolation criteria defined above.

\begin{figure}[t]
\begin{center}
  \subfigure[]{\includegraphics[width=0.485\textwidth]{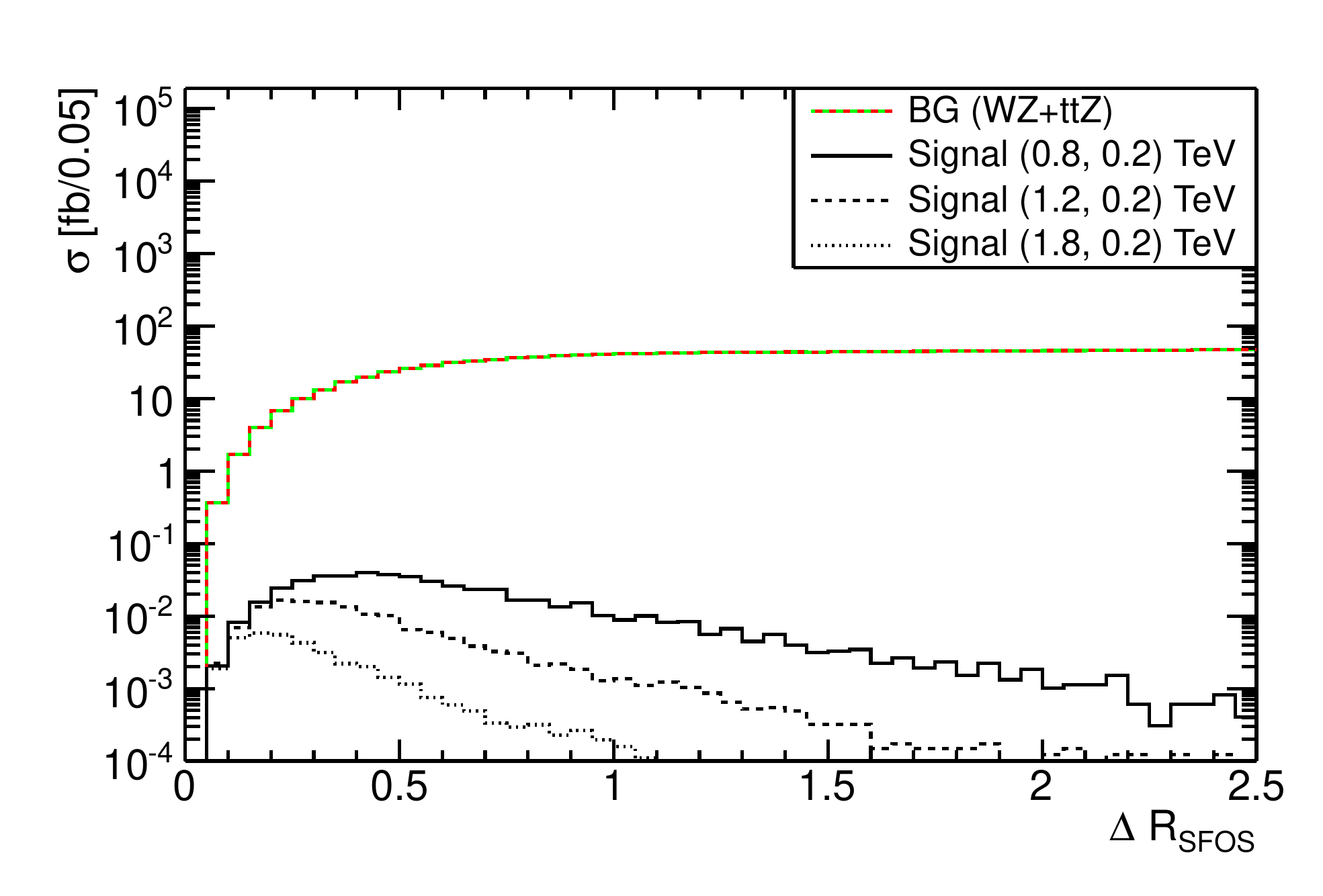}\label{fig:dRsfos-presel}} \hspace{0.1cm}
  \subfigure[]{\includegraphics[width=0.485\textwidth]{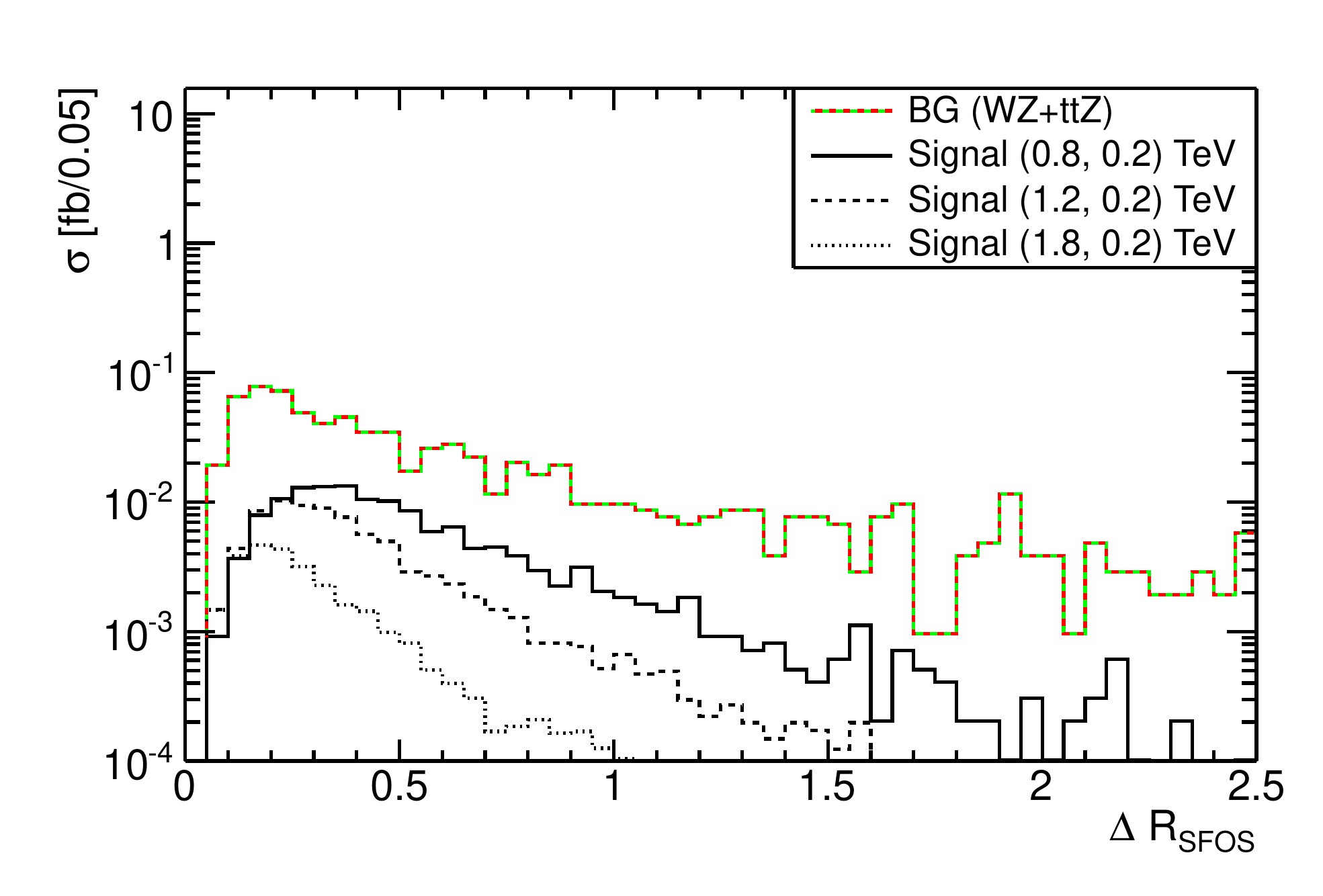}\label{fig:dRsfos-moreCuts}} 
\caption{ The distributions of $\Delta R_{\rm SFOS}$, the distance between the  SFOS lepton pair, {\bf (a)} after preselection cuts, {\bf (b)} after additional cuts: 
$\met> 500\gev$ and $m_{T}>200\gev$. 
For both plots, detector simulation has been done by {\tt Delphes 3} using the same detector setup as the one used in {\sf Snowmass} samples but with $R = 0.05$. 
\label{fig:dRsfos}
}
\end{center}
\end{figure}

To see the impact of this effect, we show the $\Delta R_{\rm SFOS}$ (the distance between the SFOS pair\footnote{
To be explicit, $\Delta R_{\rm SFOS}=\sqrt{(\Delta \phi_{\rm SFOS})^2 + (\Delta \eta_{\rm SFOS})^2}$, where $\Delta \phi_{\rm SFOS}$ and $\Delta \eta_{\rm SFOS}$
are the azimuthal and pseudo-rapidity differences between the SFOS lepton pair.
}) distributions in Fig~\ref{fig:dRsfos}.
In Fig~\ref{fig:dRsfos}, the background sample consists of the most relevant processes, $WZ$ and $ttZ$, which we have generated 
using ${\tt MadGraph~5}$ and ${\tt Phythia~6}.$\footnote{
In the $WZ$ sample, two extra partons are matched with the parton shower radiation with the MLM merging scheme \cite{Mangano:2006rw}. 
}
For signal, we examine three benchmark points:
$(M_{2},\mu)/{\rm GeV}=\ (800$, 200), (1200, 200) and (1800, 200). 
The particle level samples are passed to {\tt Delphes~3} with the same detector setup as used in {\sf Snowmass} but with $R=0.05$ for the lepton isolation cone radius.

Fig.~\ref{fig:dRsfos-presel} shows the $\Delta R_{\rm SFOS}$ distributions after the preselection cuts.
As can be seen, signal events are more concentrated around the small $\Delta R_{\rm SFOS}$ values, while background has rather flat distribution.
One can also see that smaller $\Delta R_{\rm SFOS}$ is preferred for model points with larger mass hierarchy.

In Fig.~\ref{fig:dRsfos-moreCuts} we present the same distributions of $\Delta R_{\rm SFOS}$ but with the requirement of $\met >500 \gev$ and $m_{T}>200 \gev$ on top of the preselection cuts. 
As can be seen, the distributions are more concentrated for signal and background compared to the distributions with only preselection cuts.
This is because the harsh cuts on $\met$ and $m_T$ call for large $\sqrt{\hat s}$ for the partonic collision, 
leading to more boosted $Z$ for both signal and background events.
One can see that the significant fraction of events has a SFOS lepton pair lying within $\Delta R_{\rm SFOS} < 0.3$ of each other,
and it is expected that the {\sf Snowmass} lepton isolation criteria with $R = 0.3$ would reject some fraction of signal and background events.
We therefore believe that employing smaller lepton isolation cone radius will improve the chargino-neutralino mass reach to some extent, although
a dedicated study in this direction is beyond the scope of this paper.

\section{The visible cross sections}\label{sec:cutflow}

In this section we report the visible cross sections (the cross section after cuts) for each step of the selection cuts for different processes.
Four sets of samples are considered for the SM background, which are defined in Table \ref{tab:SMBG}.
We show the results for three benchmark model points for signal: $(M_2, \mu)/{\rm GeV} = (800, 200)$, (1200, 200) and (1800, 200). 
The (visible) cross sections with k-factor = 3 are shown in fb for all tables in this section.
Table \ref{tab:preselection} shows the (visible) cross sections for the cuts employed in the $preselection$ stage.
Table \ref{tab:loose}, \ref{tab:medium} and \ref{tab:tight} show the visible cross sections for the cuts used in {\it Loose}, {\it Medium} and {\it Tight}
signal regions, respectively. 
The last columns in Tables \ref{tab:loose}, \ref{tab:medium} and \ref{tab:tight} show $S/\sqrt{B}$ assuming 3000 fb$^{-1}$ of integrated luminosity
for the three different benchmark points.

\begin{table}[h]
\vspace{5mm}
\begin{center}
\begin{tabular}{c|c|c|c|c}
\hline
Process & No cut  & $=3$ lepton & $|m^{\rm SFOS}_{\ell \ell} - m_Z| < 10$  &  no-$b$ jet \\
\hline
VV         & 3025348 &  2487
        &    2338 &
  2176        \\
\hline
ttV         & 220161                            &  792          &    552
        &  318
   \\
\hline
tV         &    2764638        &    68.9
        &    6.07       &  4.12   \\
\hline
VVV         &   36276                             &    76.1
        &    56.2
        &  56.2   \\
\hline
BG total     &  6046422                            &     3424       &   2952     &  2554   \\
\hline
\hline
$(M_2, \mu) = (800, 200)$    & 1.640  &  0.588  &  0.565  &  0.534  \\
\hline
$(M_2, \mu) = (1200, 200)$    & 0.397  &  0.124  &  0.119  &  0.111  \\
\hline
$(M_2, \mu) = (1800, 200)$    & 0.0863  &  0.0190  &  0.0179  &  0.0170  \\
\hline
\end{tabular}
\end{center}
\vspace{-3mm}
\caption{The (visible) cross sections (in fb) for the cuts employed in the {\it preselection}. 
The column marked "No cut" shows the cross sections for the background processes (defined in Table \ref{tab:SMBG}) and the cross section times branching ratio into 3 leptons via $WZ$ for signal benchmark points. 
\color{black}
%In the column marked "No cut", one can see the total cross section for background processes, and the cross section times branching ratio into 3 leptons via $WZ$ for signal benchmark points. 
}
\label{tab:preselection}
\end{table}%

\begin{table}[h]
\vspace{5mm}
\begin{center}
\begin{tabular}{c|c|c|c||c}
\hline
Process & $p_T^{\ell} > (100, 50, 10)$ & $E_T^{\rm miss} > 150$ & $m_T > 150$   &  $S/\sqrt{B}$ \\
\hline
VV         & 647 &     106     &     5.1       &     \\
\hline
ttV         &      176                &     41.2    &     6.6       &     \\
\hline
tV         &    0.665         &    0.391
        &   0.0793      &     \\
\hline
VVV         &   23.4           &   6.0
         &     1.06      &     \\
\hline
BG total     & 847                         &     153       &   12.8     &     \\
\hline
\hline
$(M_2, \mu) = (800, 200)$    & 0.506  &  0.465  &  0.381  &  5.82  \\
\hline
$(M_2, \mu) = (1200, 200)$    & 0.109  &  0.103  &  0.090  &  1.38  \\
\hline
$(M_2, \mu) = (1800, 200)$    & 0.0168  &  0.0164  &  0.0150  &  0.234  \\
\hline
\end{tabular}
\end{center}
\vspace{-3mm}
\caption{The visible cross sections (in fb) used in the {\it Loose} signal region.  The last column shows $S/\sqrt{B}$ assuming the 3000 fb$^{-1}$ luminosity for different benchmark points.}
\label{tab:loose}
\end{table}%

\begin{table}[h]
\vspace{5mm}
\begin{center}
\begin{tabular}{c|c|c|c||c}
\hline
Process & $p_T^{\ell} > (250, 150, 50)$ & $E_T^{\rm miss} > 350$ & $m_T > 300$   &  $S/\sqrt{B}$ \\
\hline
VV         & 33.8
 &    3.13
        &    0.106
        &     \\
\hline
ttV         &    9.84
                            &   0.780
         &   0.119
         &     \\
\hline
tV         &  0.037
                              &     0.0213
       &   0.00132
         &     \\
\hline
VVV         &    1.87                        &   0.291
         &   0.0442 &     \\
\hline
BG total     &    45.6                  &     4.22       &    0.271     &     \\
\hline
\hline
$(M_2, \mu) = (800, 200)$    & 0.170  &  0.107  &  0.0845  &  8.89  \\
\hline
$(M_2, \mu) = (1200, 200)$    & 0.0572  &  0.0463  &  0.0408  &  4.30  \\
\hline
$(M_2, \mu) = (1800, 200)$    & 0.0099  &  0.0088  &  0.0081  &  0.845  \\
\hline
\end{tabular}
\end{center}
\vspace{-3mm}
\caption{The visible cross sections (in fb) used in the {\it Medium} signal region.  The last column shows $S/\sqrt{B}$ assuming the 3000 fb$^{-1}$ luminosity for different benchmark points.}
\label{tab:medium}
\end{table}%

\begin{table}[h]
\vspace{5mm}
\begin{center}
\begin{tabular}{c|c|c|c||c}
\hline
Process & $p_T^{\ell} > (400, 200, 75)$ & $E_T^{\rm miss} > 800$ & $m_T > 1100$   &  $S/\sqrt{B}$ \\
\hline
VV         & 5.65 &    0.123        &    0.00166
        &     \\
\hline
ttV         &   1.03
  &   0.0056 &    0.00092        &     \\
\hline
tV         &     0.015                           &       0.0001         &     0     &   \\
\hline
VVV         &    0.350                          &     0.0109
       &   0.00153     &     \\
\hline
BG total     & 7.05      &    0.140        &     0.00411       &     \\
\hline
\hline
$(M_2, \mu) = (800, 200)$    & 0.0460  &  0.0020  &  0.0012  &  1.00  \\
\hline
$(M_2, \mu) = (1200, 200)$    & 0.0238  &  0.0070  &  0.0052  &  4.45  \\
\hline
$(M_2, \mu) = (1800, 200)$    & 0.0053  &  0.0031  &  0.0026  &  2.22  \\
\hline
\end{tabular}
\end{center}
\vspace{-3mm}
\caption{The visible cross sections (in fb) used in the {\it Tight} signal region.  The last column shows $S/\sqrt{B}$ assuming the 3000 fb$^{-1}$ luminosity for different benchmark points.}
\label{tab:tight}
\end{table}%

\newpage

\bibliographystyle{JHEP}

\bibliography{ewkinos-100tev_arxiv}

\end{document}